\documentclass[10pt,10pt]{article}
\usepackage{beraserif}
\usepackage[T1]{fontenc}
\usepackage[latin9]{inputenc}
\usepackage{geometry}
\usepackage{color}
\usepackage{amsmath}
\usepackage{amssymb}
\usepackage{graphicx}
\usepackage[numbers]{natbib}
\usepackage[unicode=true,pdfusetitle,
 bookmarks=true,bookmarksnumbered=false,bookmarksopen=false,
 breaklinks=true,pdfborder={0 0 0},backref=section,colorlinks=true]
 {hyperref}

\makeatletter

\providecommand{\tabularnewline}{\\}

\newcommand{\lyxaddress}[1]{
\par {\raggedright #1
\vspace{1.4em}
\noindent\par}
}

\makeatother

\begin{document}

\title{Solar and planetary oscillation control on climate change: hind-cast,
forecast and a comparison with the CMIP5 GCMs}

\author{Nicola Scafetta$^{1}$}

\maketitle

\lyxaddress{$^{1}$Active Cavity Radiometer Irradiance Monitor (ACRIM) Lab, Coronado,
CA 92118, USA, \& Duke University, Durham, NC 27708, USA}

\lyxaddress{email: nicola.scafetta@gmail.com}
\begin{abstract}
Global surface temperature records (e.g. HadCRUT4) since 1850 are
characterized by climatic oscillations synchronous with specific solar,
planetary and lunar harmonics superimposed on a background warming
modulation. The latter is related to a long millennial solar oscillation
and to changes in the chemical composition of the atmosphere (e.g.
aerosol and greenhouse gases). However, current general circulation
climate models, e.g. the CMIP5 GCMs, to be used in the AR5 IPCC Report
in 2013, fail to reconstruct the observed climatic oscillations. As
an alternate, an empirical model is proposed that uses: (1) a specific
set of decadal, multidecadal, secular and millennial astronomic harmonics
to simulate the observed climatic oscillations; (2) a 0.45 attenuation
of the GCM ensemble mean simulations to model the anthropogenic and
volcano forcing effects. The proposed empirical model outperforms
the GCMs by better hind-casting the observed 1850-2012 climatic patterns.
It is found that: (1) about 50-60\% of the warming observed since
1850 and since 1970 was induced by natural oscillations likely resulting
from harmonic astronomical forcings that are not yet included in the
GCMs; (2) a 2000-2040 approximately steady projected temperature;
(3) a 2000-2100 projected warming ranging between 0.3 $^{o}C$ and
1.6 $^{o}C$, which is significantly lower than the IPCC GCM ensemble
mean projected warming of 1.1 $^{o}C$ to 4.1 $^{o}C$; ; (4) an equilibrium
climate sensitivity to $CO_{2}$ doubling centered in 1.35 $^{o}C$
and varying between 0.9 $^{o}C$ and 2.0 $^{o}C$. 

\textbf{Cite as: Scafetta, N., Solar and planetary oscillation control
on climate change: hind-cast, forecast and a comparison with the CMIP5
GCMs. Energy \& Environment: special volume \textquoteleft{}Mechanisms
of Climate Change and the AGW Concept: a critical review\textquoteright{}.
Vol. 24 (3\&4), Page 455-496 (2013). DOI} \href{http://multi-science.metapress.com/content/p7n531161076t3p6/?p=80a5c828a3544b998d9a28f3fbe2baf9&pi=10}{10.1260/0958-305X.24.3-4.455} 
\end{abstract}

\section{Introduction}

Since 1850 the global surface temperature (GST) increased by 0.8-0.85
$^{o}C$, and since the 1970s by 0.5-0.55 $^{o}C$. Figure 1 depicts
the HadCRUT4 (1850-2012) GST record%
\footnote{\href{http://www.metoffice.gov.uk/hadobs/hadcrut4/}{http://www.metoffice.gov.uk/hadobs/hadcrut4/}%
} \citep{Morice}. The observed secular warming occurred during a period
of increasing atmospheric concentrations of greenhouse gases (GHG),
especially $CO_{2}$ and $CH_{4}$, likely due to human emissions
\citep{IPCC}. Current general circulation models (GCMs) interpret
that anthropogenic climatic forcings caused more than 90\% of the
global warming since 1900 and virtually 100\% of the global warming
since 1970. This hypothesis is known as the \textit{Anthropogenic
Global Warming Theory} (AGWT). Based on GCM projections, various anthropogenic
emission scenarios for the 21st century predict average warming between
1 $^{o}C$ and 4 $^{o}C$ (see Fig. 1) \citep{Knutti2012}. The \textit{Intergovernmental
Panel on Climate Change} (IPCC), sponsored by the United Nations Environment
Program (UNEP) and the World Meteorological Organization (WMO), advocates
the AGWT.

\begin{figure*}[!t]
\centering{}\includegraphics[width=1\textwidth]{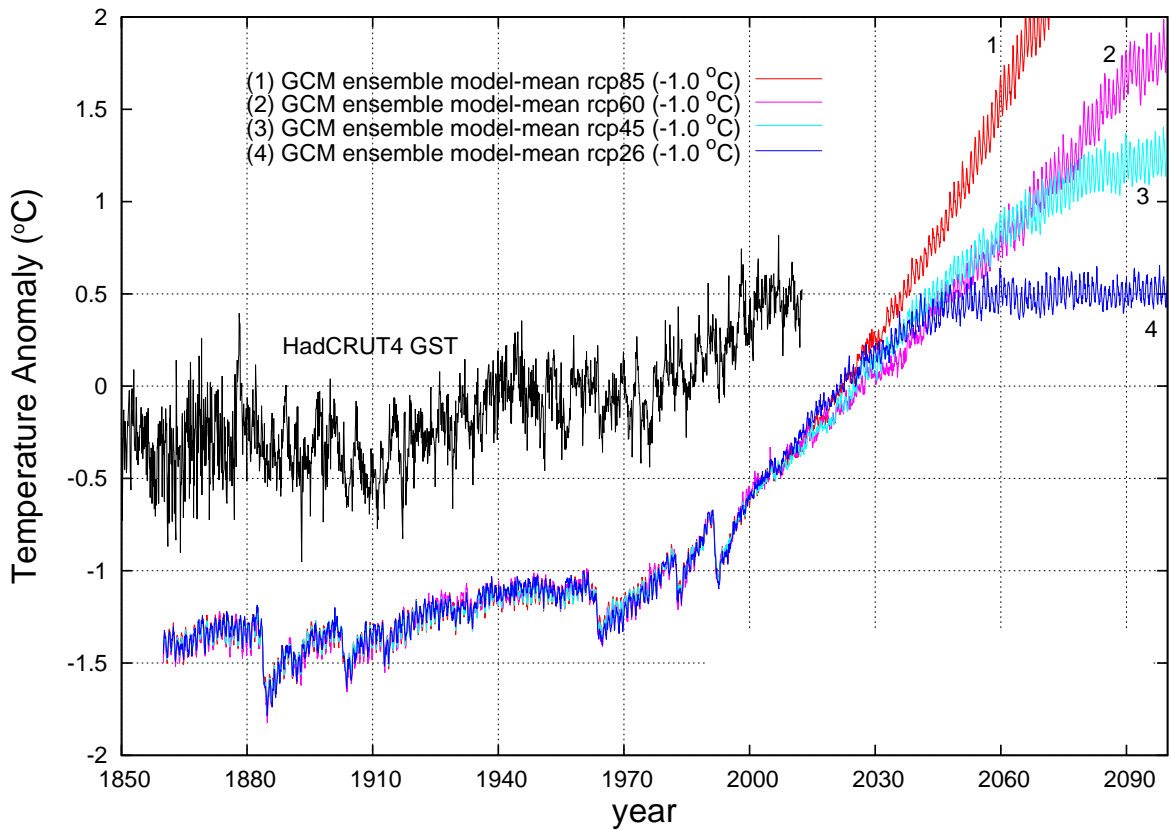}\caption{HadCRUT4 (1850-2012) GST (black) \citep{Morice}. Four Coupled Model
Intercomparison Project 5 (CMIP5) GCM ensemble mean simulations based
on known historical forcings (1860-2006) and four alternate 21st century
emission projections. The records are shifted by 1 $^{o}C$ for visualization. }
\end{figure*}

The \citet{IPCC} justified its interpretation and predictions by
the results of GCM climatic simulations as akin to those shown in
figures 9.5a and 9.5b of its AR4 report%
\footnote{\href{http://www.ipcc.ch/publications_and_data/ar4/wg1/en/figure-9-5.html}{http://www.ipcc.ch/publications\_{}and\_{}data/ar4/wg1/en/figure-9-5.html}%
}. These figures compare the GCM effects of all known natural and anthropogenic
forcings with those of natural (solar and volcano) forcings only.
It was claimed that : (1) natural forcings alone could only have induced
a negligible warming since 1900 and a slight cooling since 1970 (fig.
9.5b); (2) only the addition of anthropogenic forcings could recover
the observed warming (fig. 9.5a).

However, since 1997-1998 no detectable warming has been observed while
the GCMs predicted an average steady warming of about 2 $^{o}C/century$
(Fig. 1). This obvious divergence between data and GCM simulations
during the last 15 years was, however, ruled out with 95\% confidence
by the same AGWT advocates \citep{Knight}. Thus, the current GCMs
appear to be misleading in so far they overestimate anthropogenic
forcings while underestimating and/or ignoring some important natural
climatic mechanisms.

Indeed, large and unresolved theoretical GCM uncertainties in climate
forcing and climate sensitivity to radiative forcing exist and were
already known \citep{IPCC}; but, 13 years ago many scientists were
convinced of the reliability of the available climate models owing
to their compatibility with the \textit{hockey-stick} shaped paleoclimatic
temperature reconstructions proposed by Mann et al. from 1998 to 2004
\citep{Mann1999,Mann2003,Crowley}. However, as it will be demonstrated
in Section 3, the IPCC AGWT interpretation collapses versus novel
paleoclimatic temperature reconstructions proposed since 2005 because
these recent reconstructions reveal a three-to-four time larger preindustrial
climatic variability.

As an alternate, a novel theory proposed by \citet{Scafetta2010,Scafetta2012a,Scafetta2012b,Scafetta2012c,Scafetta2012d}
is summarized. The author found that: (1) the climate system is mostly
characterized by a specific set of oscillations; (2) these oscillations
appear to be synchronous with major astronomical oscillations (solar
system, solar activity and long solar/lunar tidal cycles); (3) these
oscillations are not reproduced by the present-day GCMs, thus indicating
that these models miss important forcings of the climate system and
related feedbacks. Therefore, an empirical model is proposed that
is based on detected decadal, multidecadal, multisecular and millennial
natural cycles plus a correction of the GCM ensemble mean simulations
to obtain an anthropogenic plus volcano climatic signature. By contrast
to the GCMs, the proposed empirical model successfully hind-casts
and reconstructs the GST patterns at multiple time scales since 1850
and approximately hind-casts general climatic patterns for centuries
and millennia. More reliable and less alarming projections for the
21st century are obtained.

\section{Unresolved physical uncertainty of current GCMs}

The reliability of the current GCMs is limited by the following five
major sources of uncertainty: 
\begin{enumerate}
\item \textsl{Climate data are characterized by various errors that can
bias composites.} For example, GST records (HadCRUT3, HadCRUT4, GISSTEM
and NCDC) present similar patterns with a net 1850-2012 warming of
about 0.8-0.85 $^{o}C$ \citep{Morice}. However, \citet{McKitrick2007}
and \citet{McKitrick2010} found that up to half of the observed 1979-2002
warming trend ($\sim$0.2 $^{o}C$) could be due to residual urban
heat island (UHI) effects, although the temperature data had already
been processed to remove the (modeled) UHI contribution. Also a \textit{divergence
problem} of proxy temperature models and instrumental records from
the 1950s onward has been observed and questions the reliability either
of the proxy models or of the instrumental GST records \citep{DArrigo,Ljungqvist}. 
\item \textsl{There may be physical processes and mechanisms that are still
unknown and, therefore, are not included in the current GCMs.} Failures
in properly modeling specific data patterns can highlight this type
of problems. If so, the limitation of the analytical GCM approach
may be partially circumvented by adopting empirical modeling that
may work well if the specific dynamics of the conjectured unknown
mechanisms are somehow identified, although the physical details of
the mechanisms themselves may remain unknown. Empirical modelling
is how ocean tides have been forecast since antiquity \citep{Ptolemy,Bede,Kepler,Kelvin,Ehret}. 
\item \textsl{The failure of GCMs may be due to not predictable chaos, internal
variability and missing forcings of the climate system.} For example,
since 2000 no warming has been observed while the IPCC GCMs predicted
on average a steady warming of about 2 $^{o}C/century$ \citep{Scafetta2012b}.
\citet{Meehl} speculated that such GST \textit{hiatus} periods could
be caused by \textit{unforced internal climatic variability} such
as occasionally deep-ocean heat uptakes. However, their adopted CCSM4
GCM did not predicted the steady temperature observed from 2000 to
2012, and produces only \textit{hiatus} periods in 2040-2050 and 2070-2080.
Essentially, because of internal dynamical chaos, it is claimed that
GCMs can only \textit{statistically,} that is in the ensemble of their
simulations, vaguely reproduce the observational data pattern means.
Alternatively, other authors postulated that the same post 2000 GCM-GST
discrepancy was the effect of small volcanic eruptions or Chinese
aerosols \citep{Kaufmann}. This interpretation was proposed despite
the fact that no increase in aerosol concentration has been observed
since 1998 \citep{Remer}. So, the issue is quite open and confused. 
\item \textsl{Radiative climate forcings used in the GCMs are characterized
by very large uncertainties.} The \citet{IPCC} (AR4 WG1 2.9.1 ``Uncertainties
in Radiative Forcing\textquotedblright{}%
\footnote{\href{http://www.ipcc.ch/publications_and_data/ar4/wg1/en/ch2s2-9-1.html}{http://www.ipcc.ch/publications\_{}and\_{}data/ar4/wg1/en/ch2s2-9-1.html}%
}) classifies the level of scientific understanding of 11 out of 16
forcing agent categories as either \textit{low} or \textit{very low}.
For example, figure SPM2%
\footnote{\href{http://www.ipcc.ch/publications_and_data/ar4/wg1/en/figure-spm-2.html}{http://www.ipcc.ch/publications\_{}and\_{}data/ar4/wg1/en/figure-spm-2.html}%
} of the \citet{IPCC} estimates a 1750-2005 net anthropogenic radiative
forcing between 0.6 and 2.4 $W/m^{2}$ and the total solar irradiance
forcing between 0.06 and 0.30 $W/m^{2}$. Given this large forcing
uncertainty, GCM modelers could arbitrarily adjust internal parameters
and forcing functions, such as the very uncertain aerosol forcing,
to improve the fit of their models to the data. Indeed, an inverse
correlation was found between the GCM modeled climate sensitivity
and total anthropogenic forcing \citep{Kiehl,Knutti}. 
\item \textsl{The current equilibrium climate sensitivity to radiative forcing
is extremely uncertain.} The \citet{IPCC} suggests that a doubling
of atmospheric $CO_{2}$ concentrations would induce a most likely
warming in the range of 2-4.5 $^{o}C$ averaging to about 3 $^{o}C$,
which is about the average value simulated by the GCMs. The total
range spans between 1-9 $^{o}C$: see Box 10.2-fig. 1 in \citet{IPCC}%
\footnote{\href{http://www.ipcc.ch/publications_and_data/ar4/wg1/en/box-10-2-figure-1.html}{http://www.ipcc.ch/publications\_{}and\_{}data/ar4/wg1/en/box-10-2-figure-1.html}%
}. In fact, while the greenhouse properties of $CO_{2}$ can be experimentally
determined (without water vapor and cloud feedbacks, doubling of $CO_{2}$
has a forcing of about 3.7 $W/m^{2}$ causing about 1 $^{o}C$ warming
\citep{Rahmstorf}), the strength of the adopted climatic feedbacks
can not be tested experimentally, and is indirectly estimated in various
ways. Some empirical studies suggest that the real climate sensitivity
may be as low as 0.5-1.3 $^{o}C$ \citep{Lindzen2011,Spencer2011}. 
\end{enumerate}
\begin{figure*}[!t]
\centering{}\includegraphics[width=1\textwidth]{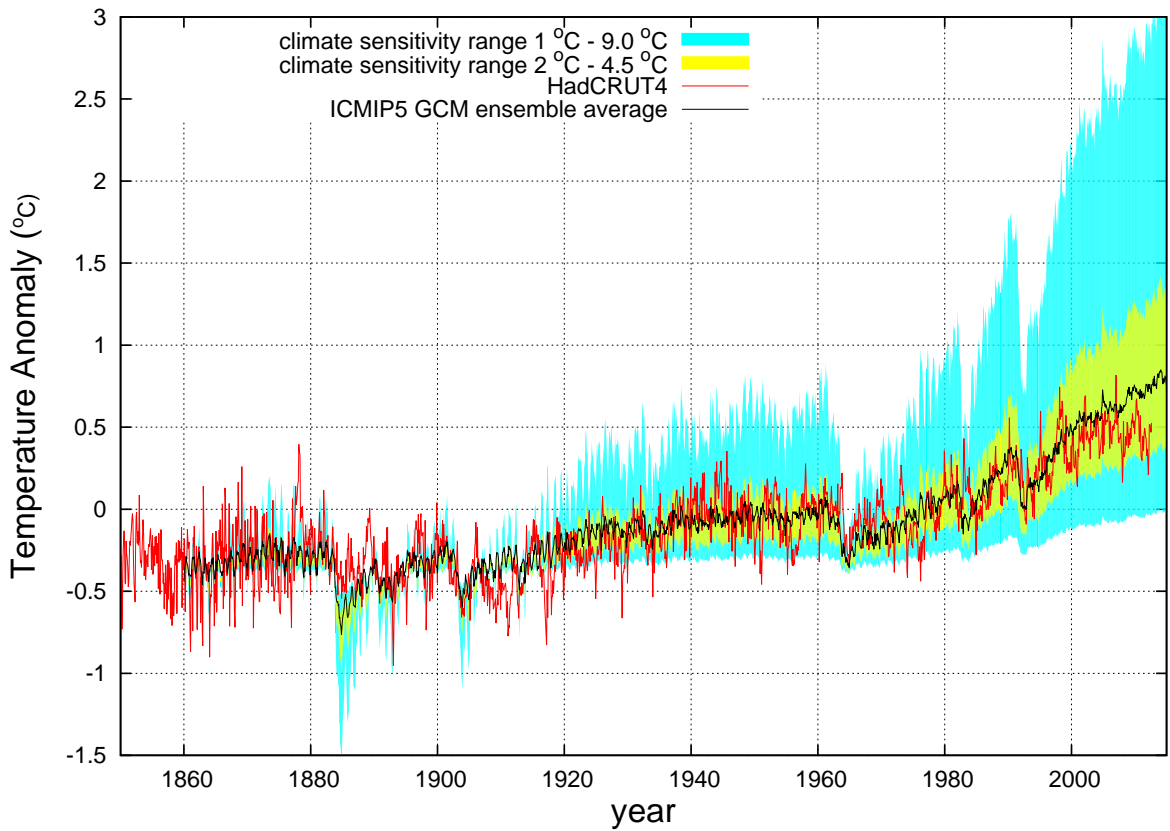}\caption{HadCRUT4 GST record (red) vs. the CMIP5 (rcp60) ensemble mean simulation
(black). Uncertainty ranges refer to equilibrium climate sensitivity
to $CO_{2}$ doubling spanning 2-4.5 $^{o}C$ (yellow), and 1-9 $^{o}C$
(cyan). }
\end{figure*}

The GCM physical uncertainties appear to be \textit{monstrous} \citep{Curry}.
It is legitimate to question whether current GCMs implement all relevant
physical mechanisms and whether their simulations and projections
can be trusted. For example, assuming an equilibrium climate sensitivity
range of 2-4.5 $^{o}C$, the net climatic forcing adopted by the GCMs
would predict a 1850-2012 net warming of about 0.57-1.3 $^{o}C$,
while with a climate sensitivity of 1-9 $^{o}C$, it would be in the
range of 0.28-2.6 $^{o}C$. As shown in Figure 2, estimated uncertainties
diverge in model predictions after 100 years progressively significantly
larger than from the data patterns that the models attempt to reconstruct.
Thus, the performance and physical reliability of these GCMs cannot
be verified within a viable accuracy while it is always possible to
adjust some model parameters or some forcing functions to obtain results
that, at a first sight, appear to reconstruct the temperature warming.

Contrary to what \citet{Knutti2012} claimed, an ensemble agreement
between different GCMs is not a guarantee of their physical reliability
implying \textit{a greater confidence in their projections} since
all models may simply reach the same erroneous conclusion by mistaking
or missing the same physical mechanisms. The scientific method requires
that comparisons must be made with observations and not only between
models. Let us see what the data tell us.

\section{AGWT agrees only with outdated hockey-stick paleoclimatic temperature
reconstructions}

Why did scientists supporting the IPCC accept the results of GCMs
and the AGWT despite the well-known large uncertainties discussed
above? This needs clarification.

In 1998-1999 \citet{Mann1999} and \citet{Mann2003} published preliminary
paleoclimatic GST reconstructions for the last 1000 years suggesting
that from the Medieval Warm Period (MWP) (900-1400) to the Little
Ice Age (LIA) (1400-1800) there was a cooling of $\sim0.2$ $^{o}C$
opposed to a drastic temperature increase of $\sim1$ $^{o}C$ since
1900. The shape of his GST resembles a \textit{hockey stick}. Despite
the fact that historically documented climate changes (e.g. the Viking
settlements in Greenland between 900 AD and 1400 AD, and many other
well-documented world-wide events \citep{Guidoboni}) contradict this
hockey-stick graph that contradicts even the IPCC First Assessment
Report (FAR, fig. 7.1, 1990)%
\footnote{\href{http://www.ipcc.ch/ipccreports/far/wg_I/ipcc_far_wg_I_chapter_07.pdf}{http://www.ipcc.ch/ipccreports/far/wg\_{}I/ipcc\_{}far\_{}wg\_{}I\_{}chapter\_{}07.pdf}%
} \citep{IPCC1990}, Mann's GST was considered trustworthy.

\begin{figure*}[!t]
\centering{}\includegraphics[width=1\textwidth,height=0.6\paperheight]{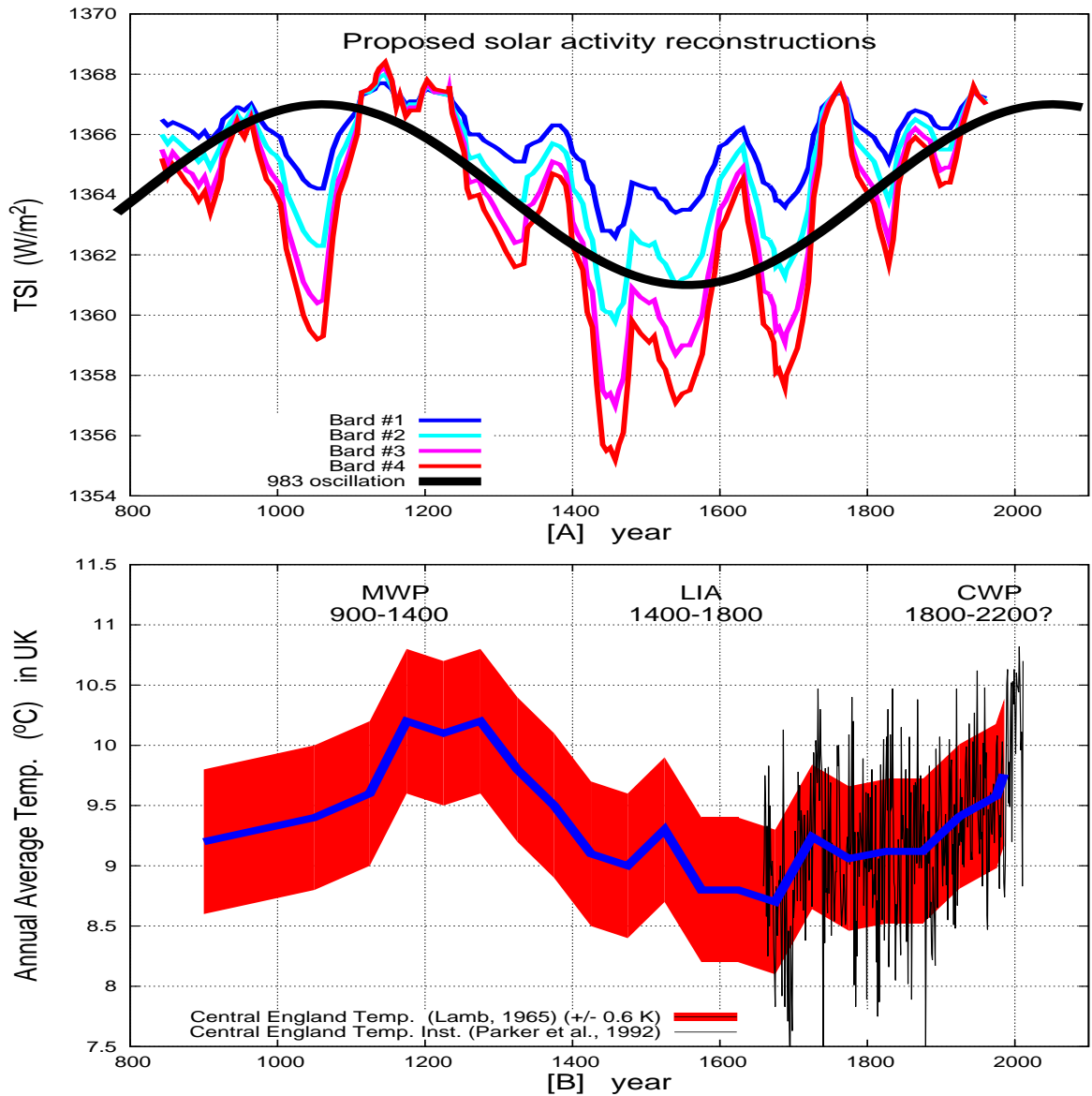}\caption{{[}A{]} Proposed solar activity reconstructions based on the solar
proxy records of $^{10}Be$ and $^{14}C$ cosmogenic isotope production
\citep{Bard}. {[}B{]} Central England: HadCET temperature record
(black) \citep{Parker} superimposed on a proxy temperature reconstruction
(red) \citep{Lamb}. }
\end{figure*}

Several groups \citep{Crowley,Hegerl,Foukal} used energy balance
models to interpret the hockey-stick temperature graphs and concluded
that the climate is poorly sensitive to solar changes and that the
post-1900 warming is almost entirely caused by anthropogenic forcing.
In 2000 \citet{Crowley} stated: \textit{The very good agreement between
models and data in the pre-anthropogenic interval also enhances confidence
in the overall ability of climate models to simulate temperature variability
on the largest scales.} Since underlying climate models were able
to hind-cast the hockey-stick proxy temperature reconstructions covering
the last 1000 years, in 2001 the \citet{IPCC2001}%
\footnote{\href{http://www.grida.no/climate/ipcc_tar/vol4/english/pdf/wg1ts.pdf}{http://www.grida.no/climate/ipcc\_{}tar/vol4/english/pdf/wg1ts.pdf}%
} could promote AGWT.

However, since 2005 a number of studies confirmed the doubts of \citet{Soon2003}
about a diffused MWP and demonstrated: (1) Mann's algorithm contained
a mathematical error that nearly always produces hockey-stick shapes
even from random data \citep{McIntyre}; (2) a global pre-industrial
temperature variability of about 0.4-1.0 $^{o}C$ between the MWP
and the LIA \citep{Ljungqvist,Moberg,Mann2008,Loehle2008,Christiansen,Esper};
(3) the existence of a millennial climatic oscillation observed throughout
the Holocene that correlates with the millennial solar oscillation
\citep{Scafetta2012c,Bond,Kerr,Ogurtsov,Kirkby,Steinhilber} and agrees
better with historical inferences \citep{Guidoboni}. Indeed, since
2001 it was clear that the climate of the last 1000 years could have
been influenced by a large millennial climatic oscillation induced
by solar activity \citep{Bond,Kerr}. Nevertheless, numerous climate
scientists claimed that the MWP affected only the North Atlantic.

For example, Figure 3B shows for Central England the HadCET instrumental
temperature record since $\sim$1700 AD \citep{Parker} and a proxy
temperature reconstruction by \citet{Lamb} since $\sim$900 AD. The
shape clearly contradicts Mann's hockey-stick GST: the impression
is that the warming trending observed since 1700 has been mostly due
to a quasi-millennial natural oscillation driven by solar activity
shown in Figure 3A \citep{Bard}. In fact, Lamb's curve suggests that
in England the MWP was as warm as, or even warmer than current temperatures.
However, findings such as Lamb-like reconstructions were dismissed
\citep{IPCC2001}. For example, \citet{Jones} claimed that Lamb's\textit{
graph was not representative of global conditions} and that \textit{the
techniques employed by Lamb were not very robust}. Nevertheless, today
such a claim needs to be questioned because recent publications support
the overall pattern of Lamb\textquoteright{}s temperature reconstruction
\citep{Ljungqvist,Loehle2008,Christiansen,Esper}. It is worth to
mention the \textit{Medieval Warm Period Project}%
\footnote{\href{http://www.co2science.org/data/mwp/mwpp.php}{http://www.co2science.org/data/mwp/mwpp.php}%
} that collects numerous peer reviewed works documenting that the MWP
was a global phenomenon.

\begin{figure*}[!t]
\centering{}\includegraphics[width=1\textwidth]{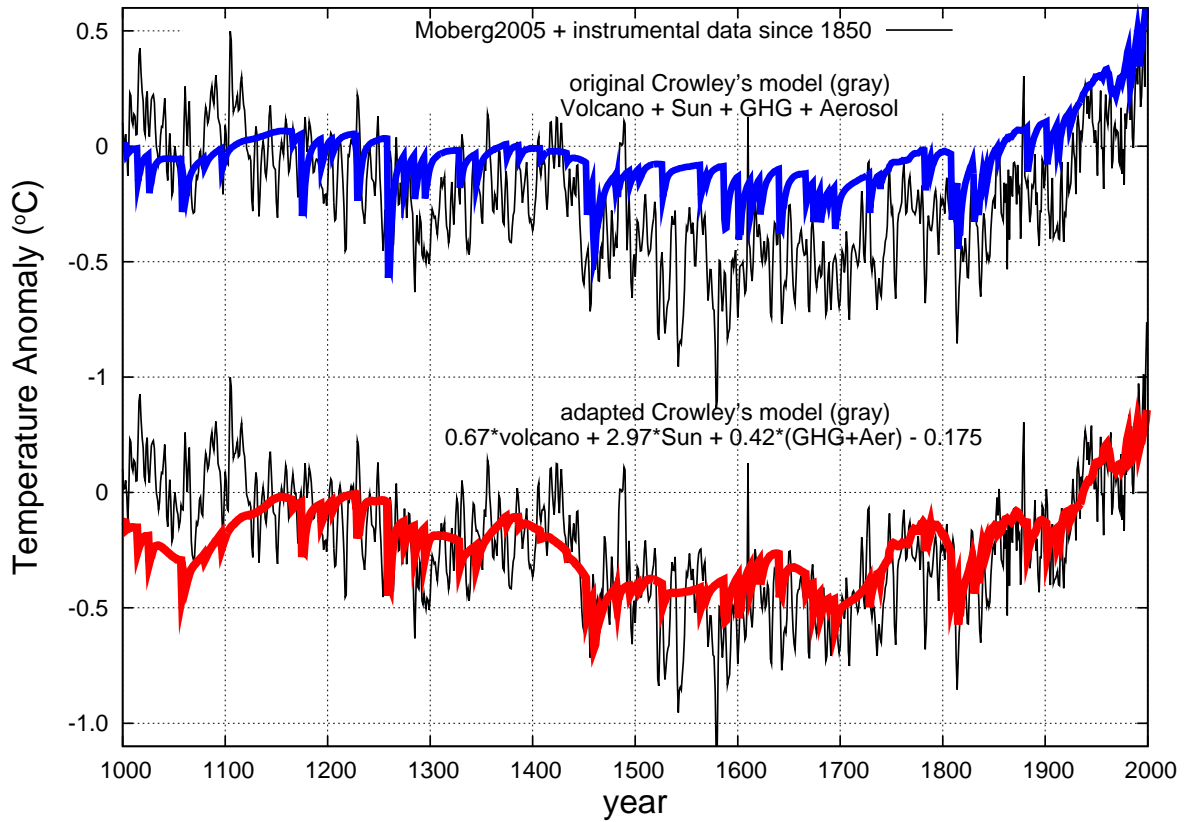}\caption{\citet{Moberg} temperature reconstruction (black) merged with the
HadCRUT4 GST after 1850. Upper panel: in blue original energy balance
model of \citet{Crowley}; bottom panel: in red empirical model integrating
volcano, solar and GHG+Aerosol temperature signature components produced
by the Crowley model rescaled to fit the Moberg temperature record.}
\end{figure*}

Figure 4 demonstrates that climate models fitting the \textit{hockey-stick}
temperature record, do not fit present-day proxy temperature reconstructions
any more. In the upper panel the original climate model of \citet{Crowley}
is superimposed on the proxy temperature model of \citet{Moberg}
(1000-1850) merged with HadCRUT4 GST (1850-2000). Their fit is poor
because Crowley's model fitted the \citet{Mann1999} hockey-stick
graph showing just a $\sim$0.2 $^{o}C$ cooling from MWP to LIA,
while Moberg GST model shows a three/four times larger cooling, $\sim$0.7
$^{o}C$, during the same period. The lower panel gives an empirical
model constructed by rescaling via linear regression Crowley's climate
model components (solar, volcano and GHG+Aerosol) for direct comparison
with Moberg GST record.

Moberg GST record implies a three times larger solar climatic impact
than the original Crowley model estimate. Its volcano effect had to
be reduced by about 30\% while the anthropogenic forcing effect (GHG
plus Aerosol forcing) by about 55\%. This implies that about 50-60\%
of the warming observed since 1900 could have been due to an increase
in solar activity that has occurred since after the 17th century Maunder
solar minimum. This result confirms \citet{Scafetta2007} and \citet{Scafetta2009},
and strongly contradicts the IPCC AR3 and AR4 \citep{IPCC,IPCC2001},
\citet{Benestad} and \citet{Lean2008} who claimed that more than
90-93\% of the 20th century warming was caused by anthropogenic GHG
emissions.

Thus, the energy balance model of \citet{Crowley} is unable to reproduce
the empirically determined solar signature evident in modern paleoclimatic
GST reconstructions, and considerably overestimates the anthropogenic
component. Today, the problem is even more significant because the
most recent GCMs (the CMIP5) use a solar forcing based on the total
solar irradiance (TSI) reconstruction of Lean \citep{Wang}, which
shows a 50\% smaller secular and millennial solar variability than
the solar model used by Crowley, which used the model by \citet{Bard}
rescaled on an earlier TSI model by \citet{Lean1995}. Therefore,
or current GCMs use severely wrong TSI forcing (see Section 9), or
they miss other solar related forcing mechanisms (e.g. chemical-based
UV irradiance-related forcing of the stratospheric temperatures and
a solar wind/cosmic ray forcing of the cloud systems \citep{Kirkby}),
or both.

Had in 2000 the current paleoclimatic temperature reconstructions
been available, Crowley and other scientists of the time would have
probably had a significantly lower confidence \textit{in the overall
ability of climate models to simulate temperature variability,} and
would not have thought that the science was sufficiently \textit{settled}.
Very likely, those same scientists would have concluded that important
climate-change mechanisms were still unknown, and needed to be researched
before they could be implemented to make reliable climate models.

Today, the AGWT \textit{consensus} appears to be an \textit{accident
of history} promoted since 2001 by the discredited hockey-stick GST
records and by the IPCC in a quite questionable way \citep{Laframboise,Montford}
and by numerous scientific organizations, such as those that in 2005
signed the Joint Science Academies statement (2005),%
\footnote{\href{http://nationalacademies.org/onpi/06072005.pdf}{http://nationalacademies.org/onpi/06072005.pdf}%
} that hastily advocated AGWT despite the scientific complexity of
the climate system and the large known uncertainties demanded prudence.
During the last decade there has been also a politically motivated
\textit{consensus seeking process} \citep{CurryWebster2012}, which
is inconclusive in questions of science,%
\footnote{It is worth reminding the famous quote attributed to Galileo Galilei:
``In questions of science, the authority of a thousand is not worth
the humble reasoning of a single individual.\textquotedblright{} %
} that has likely interfered with the acquisition and interpretation
of evidences by discriminating opinions critical of the AGWT at major
science journals \citep{Lindzen}. This had also the effect to generate
a serious \textit{tension} between the AGWT advocates \citep{Gleick}
and the critical voices\textit{.} However, because numerous evidences
contradict the hockey-stick GST graph used since 2001 by the IPCC
to promote the AGWT and the GST stopped to rise 15 years ago contrary
to all GCM predictions \citep{Scafetta2012b} (Fig. 1), today a careful
investigation on the climate change attribution problem is necessary
and legitimate.

\begin{figure*}[!t]
\centering{}\includegraphics[width=1\textwidth,height=0.6\paperheight]{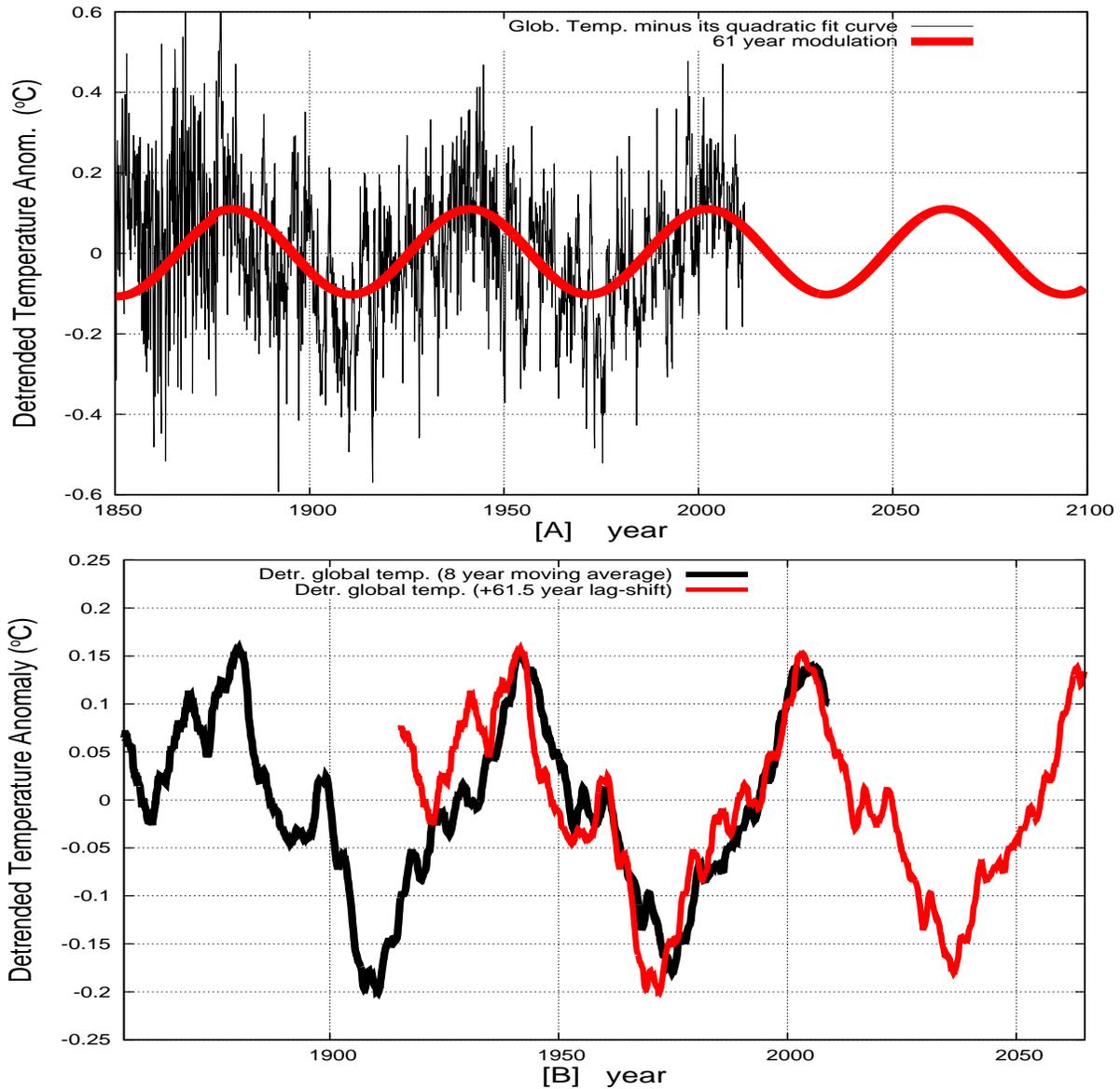}\caption{{[}A{]} HadCRUT4 GST record after detrending of its upward quadratic
trend showing a quasi 61-year modulation. {[}B{]} 8-year moving average
of the detrended GST plotted against itself with a 61.5-year shift
(red). The quadratic fitting trend applied is $f(t)=0.0000297*(t-1850)^{2}-0.384$.
For details see \citet{Scafetta2010}. }
\end{figure*}

\section{Decadal and multidecadal climatic oscillations are synchronous to
major astronomical cycles}

Geophysical systems are characterized by oscillations at multiple
time scales from a few hours to hundred thousands and millions of
years \citep{House}. Quasi decadal, bidecadal, 60 year , 80-90 year
, 115 year , 1000 year and other oscillations are found in global
and regional temperature records, in the Atlantic Multidecadal Oscillation
(AMO), North Atlantic Oscillation (NAO) and Pacific Decadal Oscillation
(PDO), in global sea level rise indexes, monsoon records, and similar
oscillations are found also in solar proxy records and in historical
aurora records covering centuries and millennia \citep[e.g.: ][]{Scafetta2010,Scafetta2012a,Scafetta2012c,Bond,Ogurtsov,Steinhilber,Agnihotri,Chylek,Cook,Currie,Davis,Gray,Hoyt,Humlum,Klyashtorin,Knudsen,Kobashi,Jevrejeva,Chambers,Mazzarella,Qian,Schulz,Sinha,Stockton,Yadava,Scafetta20131}.

\begin{figure*}[!t]
\centering{}\includegraphics[width=1\textwidth]{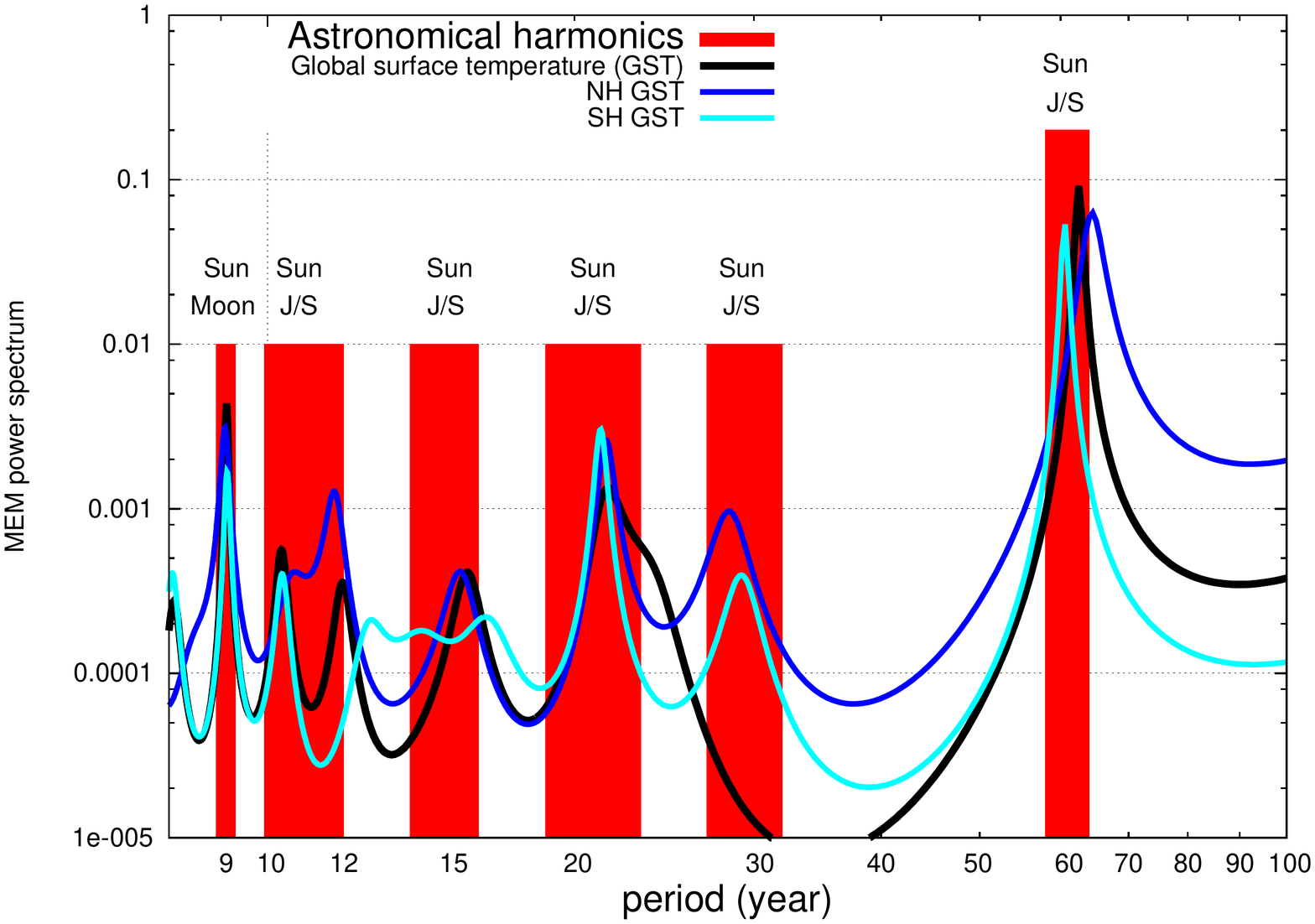}\caption{Maximum entropy method (MEM) power spectrum evaluations \citep{Courtillot}
of HadCRUT4 GST for the Northern Hemisphere GST and Southern Hemisphere
GST. The red bars represent the major expected astronomical oscillations
due to soli-lunar tidal cycles (9.1 year), and to solar cycles and
gravitational oscillations of the heliosphere due to Jupiter and Saturn
(after \citet{Scafetta2010}).}
\end{figure*}

Figures 1 and 5 show that the temperature oscillates with a quasi
61-year cycle superimposed to a general warming trend. We observe
the following 30-year periods of warming 1850-1880, 1910-1940, 1970-2000;
and the following periods of cooling 1880-1910, 1940-1970, 2000-2030(?).
By detrending the long-term warming trend,%
\footnote{This is done with a quadratic function, which captures the observed
warming acceleration and is as orthogonal as possible to the multidecadal
oscillations observed in the data.%
} the quasi 61-year oscillation can be highlighted, as shown in Figure
5 \citep{Scafetta2010}, where an almost perfect match between the
1880-1940 and 1940-2000 GST periods emerges.

\begin{figure*}[!t]
\centering{}\includegraphics[width=1\textwidth,height=0.6\paperheight]{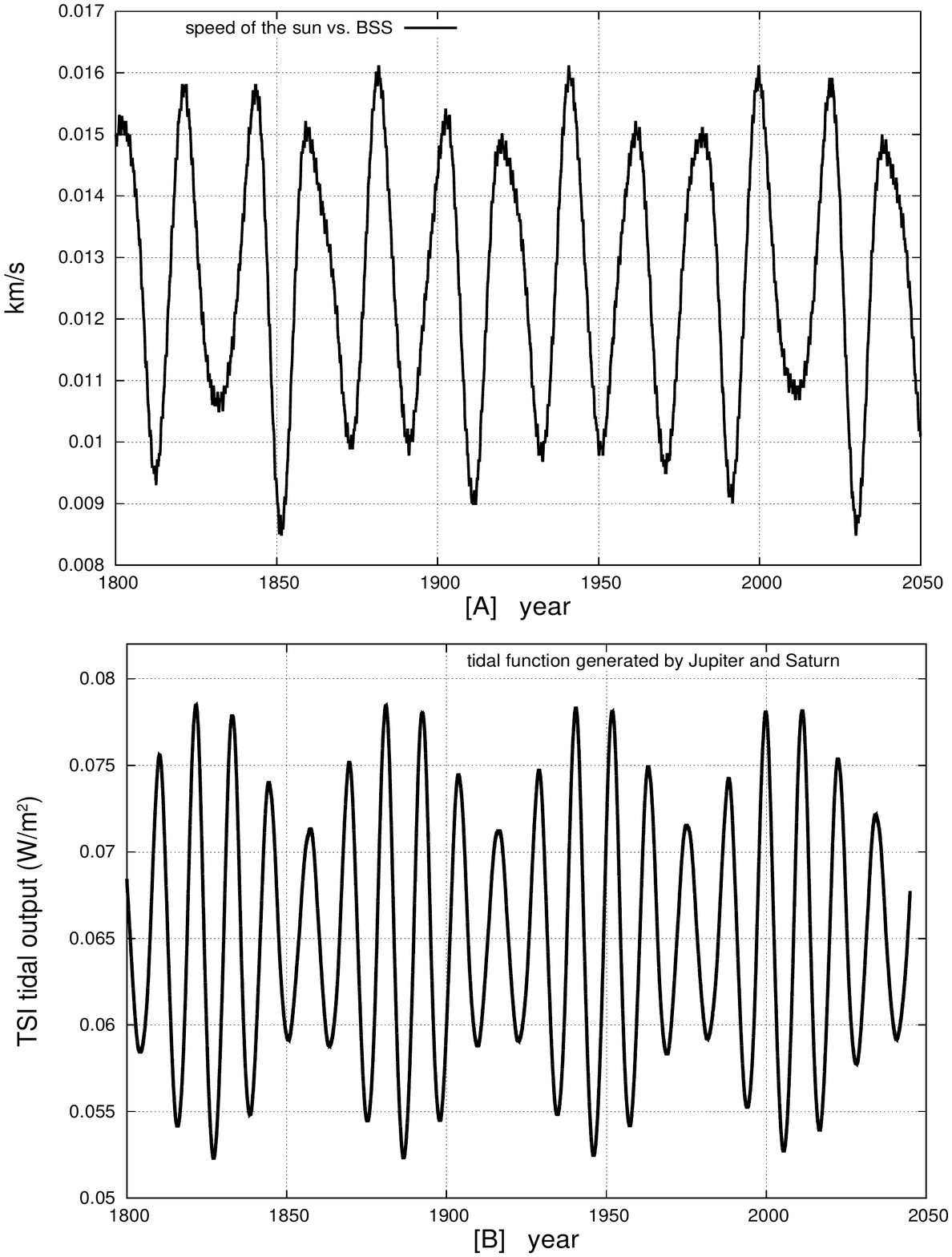}\caption{{[}A{]} Wobbling speed of the Sun \citep{Scafetta2010}. {[}B{]} Estimate
of total solar irradiance induced by Jupiter and Saturn tides. Note
the 10-12 year, 20 year and 60-61 year oscillations (for details see
\citet{Scafetta2012c,Scafetta2012d}). }
\end{figure*}

Figure 6 gives power spectrum evaluations for the novel HadCRUT4 GST,
and for the GST of the Northern (NH) and Southern hemispheres (SH)
\citep{Morice}. There are two major multidecadal oscillations with
approximate periods of 19-23 years and 59-63 years, plus two decadal
oscillations at about 8.9-9.3 years and 10-12 years. Figure 6 also
demonstrates that both hemispheres are characterized by a synchronized
climate because they present a similar set of spectral peaks \citep[see also: ][]{Scafetta2010,Scafetta2012b}.
\citet{Scafetta2010,Scafetta2012a,Scafetta2012b,Scafetta2012c,Scafetta2012d}
noted that the GST oscillations appear to be synchronized with astronomical
oscillations, which are highlighted as red boxes in Figure 6.

The 9.1 year oscillation probably relates to a major soli-lunar gravitational
tidal cycle \citep{Scafetta2010,Scafetta2012b,Wang2012}. In fact,
the lunar nodes complete a revolution in 18.6 years, and the Saros
soli-lunar eclipse cycle completes a revolution in 18 years and 11
days. These two cycles induce 9.3 year and 9.015 year tidal oscillations
corresponding respectively to Sun-Earth-Moon and Sun-Moon-Earth tidal
configurations. Moreover, the lunar apsidal precession completes one
rotation in 8.85 years causing a corresponding lunar tidal cycle.
Thus, there are three interfering major tidal cycles clustered between
8.85 year and 9.3 year periods, which generate a major oscillation
with an average period of about 9.06 years. \citet[supplement pp. 35-36]{Scafetta2012b}
showed that in 1997-1998 and 2006-2007 eclipses occurred close to
the March and September equinoxes, that is when the soli-lunar spring
tidal bulge peaks on the equator, having the strongest torquing effect
on the ocean. Filtering methodologies showed the $\sim$9.1 year GST
cycle to peak in 1997-1998 and 2006-2007 as expected \citep{Scafetta2012b}.
The Moon also causes an 18.6 year nutation cycle of the Earth's axis,
which may contribute to an 18.6 year climate oscillation \citep{Currie}.
This 18.6 year oscillation presumably interferes with the two bi-decadal
cycles of solar/planetary origin (discussed below), thus contributing
to modulate a bidecadal cycle with an average period varying between
18 and 23 years. Other long soli-lunar tidal oscillations may exist.
The solar system is also characterized by a set of natural harmonics
associated with solar cycles (e.g. the $\sim$11-year Schwabe sunspot
cycle and the $\sim$22-year Hale magnetic cycle \citep{Hoyt}) and
planetary harmonics: see Section 7.

\begin{figure*}[!t]
\centering{}\includegraphics[width=1\textwidth,height=0.6\paperheight]{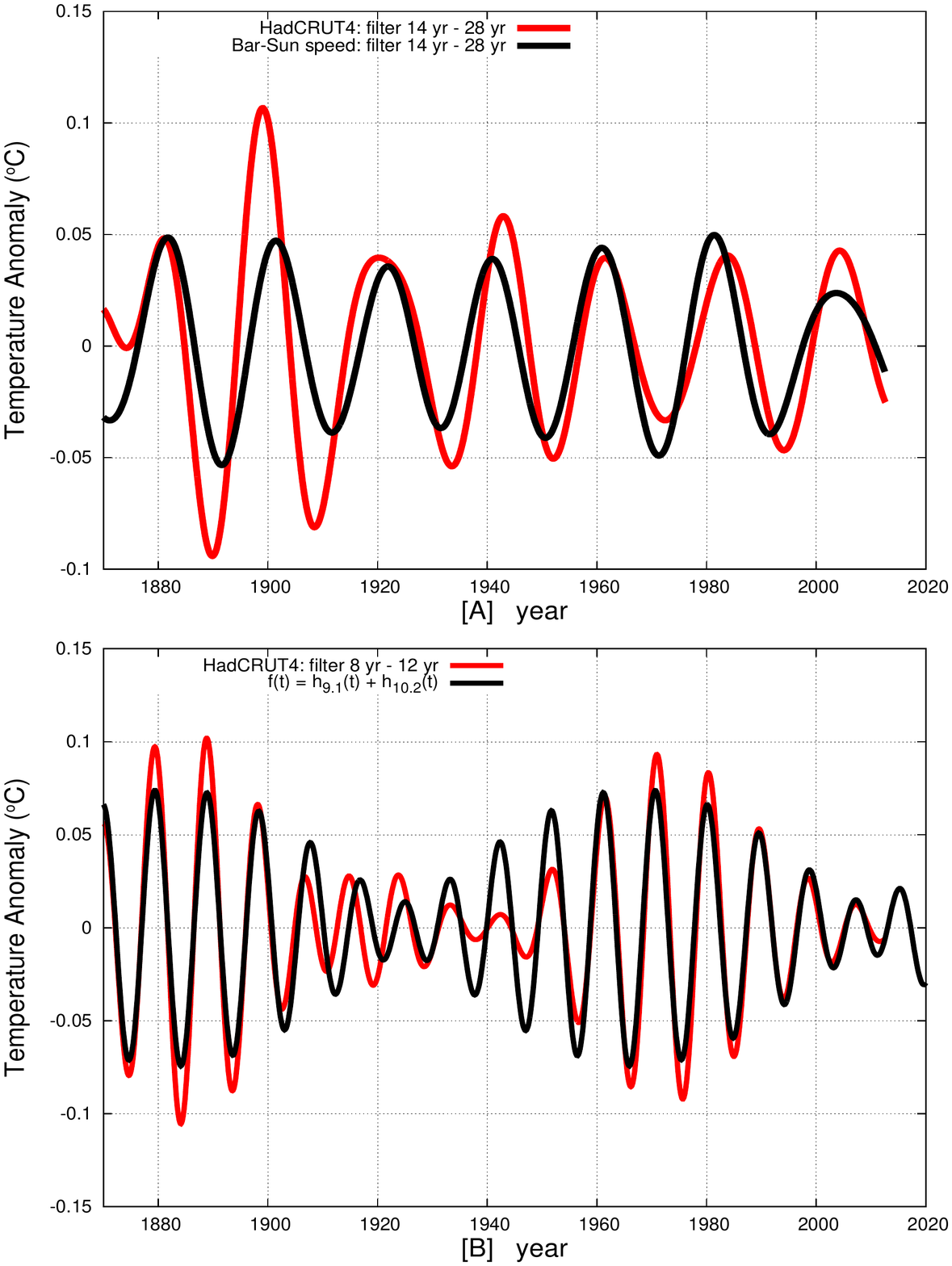}\caption{{[}A{]} Fourier filtering within 14-28 year of the HadCRUT4 GST and
SWS given in Figure 7A. {[}B{]} Fourier filtering of GST within the
period band between 8 years and 12 years compared with the regression
model from 1870 to 2012, $f(t)=h_{9.1}(t)+h_{10.2}(t)$. }
\end{figure*}

Indeed, decadal and multidecadal oscillations are clearly reflected
by the speed of the wobbling Sun (SWS) relative to the barycenter
of the solar system \citep{Scafetta2010,Scafetta2012a}. Figure 7A
shows a clear quasi 20-year oscillation and a slight increase of the
solar speed every 60 years. This occurred around 1880, 1940 and 2000,
when GST maxima occurred (Fig. 5). In addition, Jupiter and Saturn
also produce specific tidal cycles on the Sun at 9.93 years (spring
tide), 11.86 years (Jupiter orbital tide), 14.97 years (minor beat
cycle between Jupiter-Saturn spring tide and Saturn orbital tide),
29.46 years (Saturn orbital tide), 60.9 years (major beat cycle between
Jupiter-Saturn spring tide and Jupiter orbital tide). Figure 7B shows
that tidal beat maxima occurred around 1880, 1940 and 2000 during
GST maxima \citep{Scafetta2012d}. As better explained in Section
7, the Schwabe 11-year sunspot cycles vary between about 9 and 13
years and are essentially constrained by the oscillations generated
by Jupiter-Saturn spring tide and Jupiter orbital tide \citep{Scafetta2012c,Scafetta2012d}.

\begin{figure*}[!t]
\centering{}\includegraphics[width=1\textwidth]{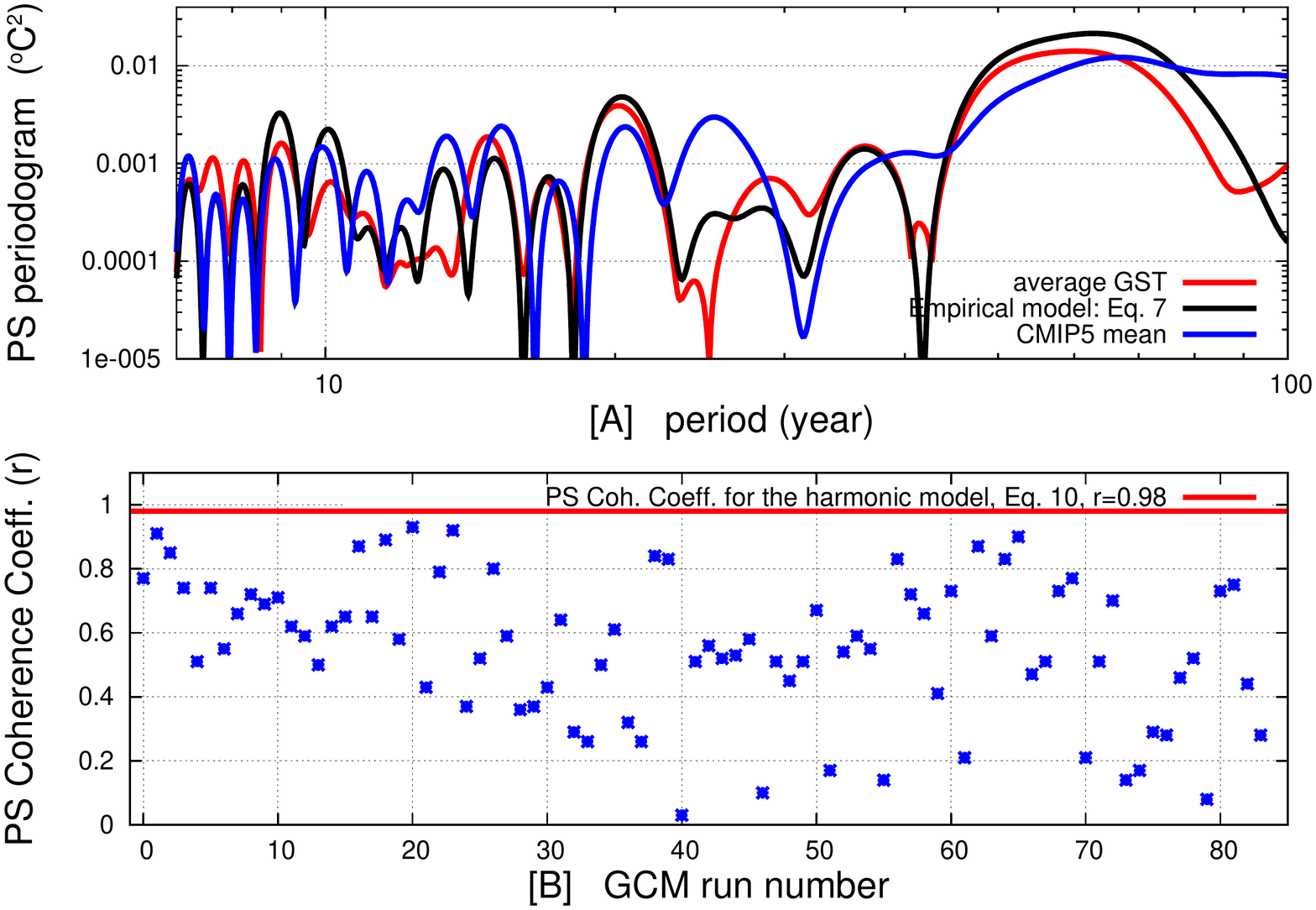}\caption{{[}A{]} Power spectrum periodograms of the GST (red), of the CMIP5
ensemble mean (blue) and of the astronomical model Eq. 7 (black).
{[}B{]} Power spectrum coherence coefficient between GST and the CMIP5
GCM runs (blue dots) as listed in Table 2. }
\end{figure*}

The decadal and multidecadal astronomical periods are given as red-boxes
in Figure 6 and correspond to the power spectral peaks observed in
the GST records, as already demonstrated by \citet{Scafetta2010}.
Figure 8A shows for the GST and SWS (Fig. 7A) a Fourier filtering
within the period band of 14-28 year demonstrating a good phase matching
of both curves. Figure 8B shows a Fourier filtering of GST within
the period band of 8-12 years, which is reconstructed with two optimal
harmonics with a 9.1 year and 10.2 year period, respectively. A 10.2
year period was used because it emerges as the main decadal peak in
Figure 6 and falls within the range referring to the 9.93-year Jupiter-Saturn
spring tide and the average 11-year solar cycle, and probably represents
part of the 11-year solar cycle effect on climate (see also \citep{Scafetta2010,Scafetta2012a,Scafetta2012b}).

GST also presents a smaller spectral peak at about 12 year, which
may be related to the Jupiter orbit: for simplicity this additional
harmonic as well as the 15-16 year and 30 year harmonics are here
ignored. The 9.1 harmonic peaked in 1997.8, as confirmed by the soli-lunar
tidal interpretation paradigm, and the other decadal cycle peaked
in 2001.5 during the end of the maximum of solar cycle 23 \citep{ScafettaWillson2009}.
Note that both a frequency and phase matching, as shown in Figures
5-8, is very important for identifying these oscillations as astronomically
induced.

By adopting the following four major constituent climatic oscillation,
regression against GST permits to obtain average optimal empirical
harmonics:

\begin{equation}
h_{9.1}(t)=0.044\cdot\cos(2\pi(t-1997.8)/9.1)\label{eq:1}
\end{equation}

\begin{equation}
h_{10.2}(t)=0.030\cdot\cos(2\pi(t-2001.5)/10.2)\label{eq:2}
\end{equation}

\begin{equation}
h_{21}(t)=0.051\cdot\cos(2\pi(t-2004.7)/21)\label{eq:3}
\end{equation}

\begin{equation}
h_{61}(t)=0.107\cdot\cos(2\pi(t-2003.14)/61)\label{eq:4}
\end{equation}
There are at least 6 major 8.85-12 year astronomic harmonics and at
least 3 major 18-23 year astronomic harmonics. Moreover, the climate
system oscillates chaotically around the signal produced by such complex
harmonic forcing function. This issue is here not further addressed
because we use a simplified model.

\begin{figure*}[!t]
\centering{}\includegraphics[width=1\textwidth]{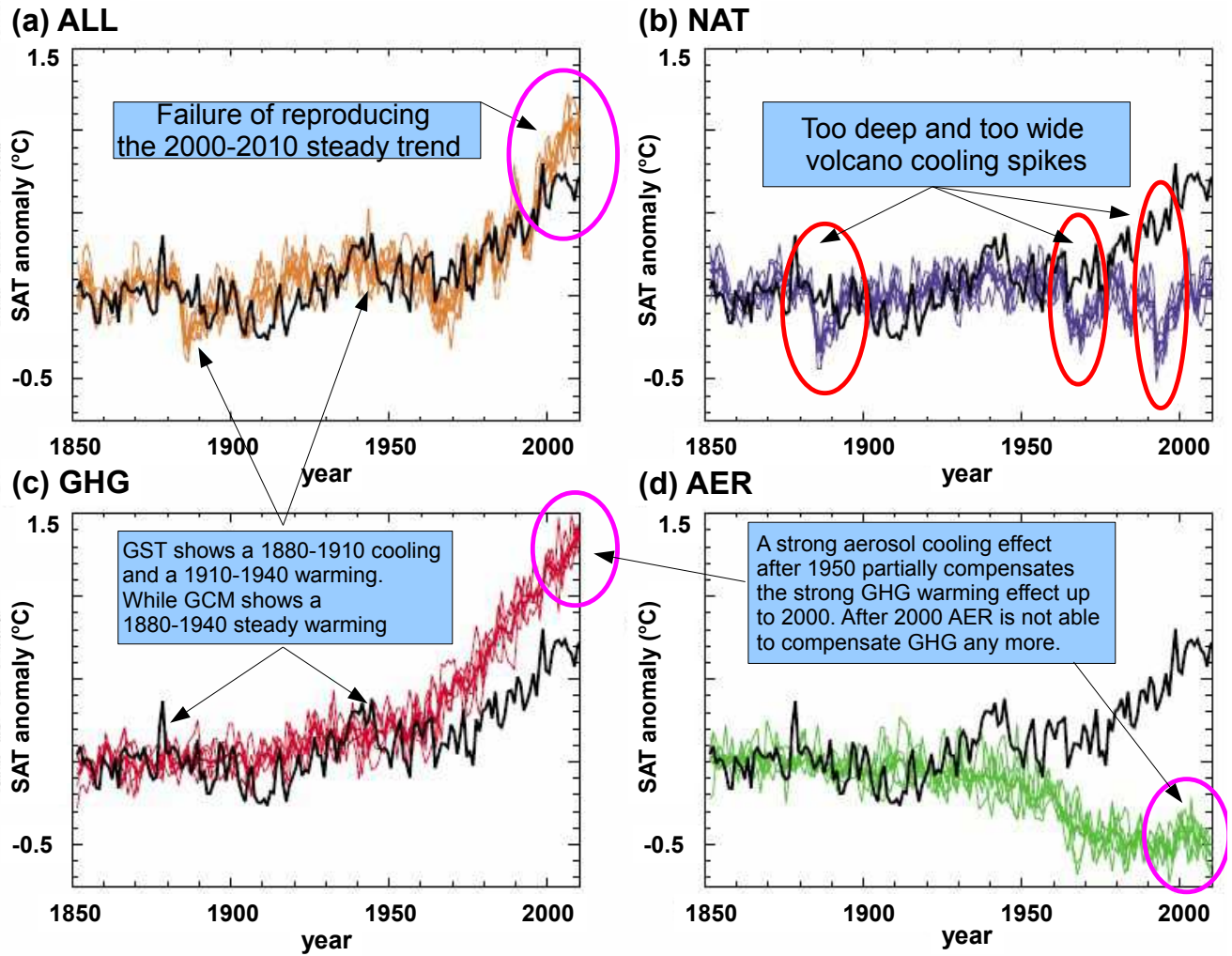}\caption{A reproduction of figure 1 of \citet{Gillett}. Comments in the diagrams
highlight common problems inherent to all CMIP5 GCMs. The computer
simulations were run with: (a) anthropogenic and natural forcings
(ALL), (b) natural forcings only (NAT), (c) greenhouse gases only
(GHG), and (d) aerosols only (AER).}
\end{figure*}

\section{CMIP3 and CMIP5 GCMs do not reconstruct the observed GST decadal
and multidecadal oscillations}

\citet{Scafetta2012b} analyzed all CMIP3 GCMs used by the \citet{IPCC}
and their individual runs, and concluded that these models do not
reproduce the decadal and multidecadal oscillations found in the GST
records. Here the 83 individual runs of the 18 CMIP5%
\footnote{KNMI Climate Explorer: \href{http://climexp.knmi.nl}{http://climexp.knmi.nl}%
} GCMs that will be used in the IPCC AR5 in 2013 are briefly subjected
to an equivalent test.

\begin{table}[!t]
\centering{}%
\begin{tabular}{|c|c|c|}
\hline 
period  & GST-trend  & GCM-trend\tabularnewline
\hline 
\hline 
1860-1880  & +1.11$\pm$0.24  & +0.54$\pm$0.06\tabularnewline
\hline 
1880-1910  & -0.57$\pm$0.09  & +0.23$\pm$0.07\tabularnewline
\hline 
1910-1940  & +1.34$\pm$0.08  & +0.90$\pm$0.03\tabularnewline
\hline 
1940-1970  & -0.27$\pm$0.09  & -0.47$\pm$0.04\tabularnewline
\hline 
1970-2000  & +1.68$\pm$0.08  & +1.66$\pm$0.05\tabularnewline
\hline 
2000-2012  & +0.40$\pm$0.25  & +1.96$\pm$0.07\tabularnewline
\hline 
\end{tabular}\caption{Comparison of 30-year period trends in $^{o}C/century$ between the
HadCRUT4 GST and the CMIP5 GCM ensemble mean simulation as given in
Figure 1.}
\end{table}
\begin{table*}[!t]
\begin{tabular}{|l|c||l|c||l|c|}
\hline 
\#-n model  & $r$  & \#-n model  & $r$  & \#-n model  & $r$\tabularnewline
\hline 
\hline 
0 model mean  & 0.77  &  &  &  & \tabularnewline
\hline 
1-0 bcc-csm 1-1  & 0.91  & 29-5  & 0.37  & 57  & 0.72\tabularnewline
\hline 
2-1  & 0.85  & 30-6  & 0.43  & 58-0 HadGEM2-ES  & 0.66\tabularnewline
\hline 
3-2  & 0.74  & 31-7  & 0.64  & 59-1  & 0.41\tabularnewline
\hline 
4-0 CanESM2  & 0.51  & 32-8  & 0.29  & 60-2  & 0.73\tabularnewline
\hline 
5-1  & 0.74  & 33-9  & 0.26  & 61-3  & 0.21\tabularnewline
\hline 
6-2  & 0.55  & 34-0 EC-Earth23  & 0.50  & 62-0 INMCM4  & 0.87\tabularnewline
\hline 
7-3  & 0.66  & 35-1  & 0.61  & 63-0 IPSL-CM5A-LR  & 0.59\tabularnewline
\hline 
8-4  & 0.72  & 36-2  & 0.32  & 64-1  & 0.83\tabularnewline
\hline 
9-0 CCSM4  & 0.69  & 37-3  & 0.26  & 65-2  & 0.90\tabularnewline
\hline 
10-1  & 0.71  & 38-4  & 0.84  & 66-3  & 0.47\tabularnewline
\hline 
11-2  & 0.62  & 39-5  & 0.83  & 67-4  & 0.51\tabularnewline
\hline 
12-3  & 0.59  & 40-6  & 0.03  & 68-0 MIROC5  & 0.73\tabularnewline
\hline 
13-4  & 0.50  & 41-0 GISS-H2-H  & 0.51  & 69-0 MIROC-ESM  & 0.77\tabularnewline
\hline 
14-5  & 0.62  & 42-1  & 0.56  & 70-1  & 0.21\tabularnewline
\hline 
15-0 CNRM-CM5  & 0.65  & 43-2  & 0.52  & 71-2  & 0.51\tabularnewline
\hline 
16-1  & 0.87  & 44-3  & 0.53  & 72-0 MIROC-ESM-CHEM  & 0.70\tabularnewline
\hline 
17-2  & 0.65  & 45-4  & 0.58  & 73-0 MPI-ESM-LR  & 0.14\tabularnewline
\hline 
18-3  & 0.89  & 46-0 GISS-E2-R  & 0.10  & 74-1  & 0.17\tabularnewline
\hline 
19-4  & 0.58  & 47-1  & 0.51  & 75-2  & 0.29\tabularnewline
\hline 
20-5  & 0.93  & 48-2  & 0.45  & 76-0 MRI-CGCM3  & 0.28\tabularnewline
\hline 
21-6  & 0.43  & 49-3  & 0.51  & 77-1  & 0.46\tabularnewline
\hline 
22-7  & 0.79  & 50-4  & 0.67  & 78-2  & 0.52\tabularnewline
\hline 
23-8  & 0.92  & 51-5  & 0.17  & 79-3  & 0.08\tabularnewline
\hline 
24-0 CSIRO-Mk3-6-0  & 0.37  & 52-6  & 0.54  & 80-4  & 0.73\tabularnewline
\hline 
25-1  & 0.52  & 53-7  & 0.59  & 81-0 NorESM1-M  & 0.75\tabularnewline
\hline 
26-2  & 0.80  & 54-8  & 0.55  & 82-1  & 0.44\tabularnewline
\hline 
27-3  & 0.59  & 55-9  & 0.14  & 83-2  & 0.28\tabularnewline
\hline 
28-4  & 0.36  & 56-0 HadGEM2-CC  & 0.83  & average  & 0.55 $\pm$0.22\tabularnewline
\hline 
\end{tabular}\caption{Correlation coefficients $r$ between the periodogram of GST and 83
individual runs for 18 GCMs. See also Figure 10B.}
\end{table*}

Figure 1 clearly shows that the CMIP5 GCM ensemble mean simulations
do not reconstruct the quasi 60-year GST oscillation observed since
1850. Table 1 summarizes 30-year trends and highlights that the GCM
ensemble mean simulations fit the GST only between 1970 to 2000, which
is just 18\% of the 162-year available period. Thus, the CMIP5 GCM
ensemble means can neither hind-cast nor forecast climate change with
a reasonable accuracy.

To test whether the CMIP5 GCMs reproduce GST oscillations, geometrical
averages were calculated for four periodograms on the base of the
HadCRUT3, HadCRUT4, GISS and NCDC GST records. Then, a periodogram
was calculated for each of the 83 individual CMIP5 GCM runs. Finally,
the correlation coefficient $r$ between the two curves was estimated
for 7 year up to 100 year periods. Only data for the common 1880-2006
interval were used and each record was linearly detrended before calculating
its periodogram. Results are summarized in Figure 9 and Table 2.

Figure 9A gives the GST periodogram (red) and the GCM ensemble mean
periodogram (blue). Figure 9B gives the correlation coefficient between
the GST periodogram and the periodogram for 83 GCM runs (blue dots).
For the GCM ensemble mean (numbered as -0-) we find $r=0.77$, while
for the other 83 GCM runs $r$ varies between 0.03 and 0.92. The average
is $\left\langle r\right\rangle =0.55\pm0.22$, and suggests that
the GCMs perform poorly in reconstructing the GST spectral characteristics.

The periodogram correlation coefficient for the astronomically-based
empirical model (see section 7) and the GST is $r=0.98$ of a possible
maximum of $r=1$. The likelihood to find an individual GCM simulation
that performs equally or better than the empirical model is less than
1\%. This finding is important as it demonstrates that the internal
variability of CMIP5 GCMs is unable to reproduce the observed GST
frequencies, suggesting that these models miss certain harmonic forcings.
\begin{figure*}[!t]
\centering{}\includegraphics[width=1\textwidth,height=0.38\textheight]{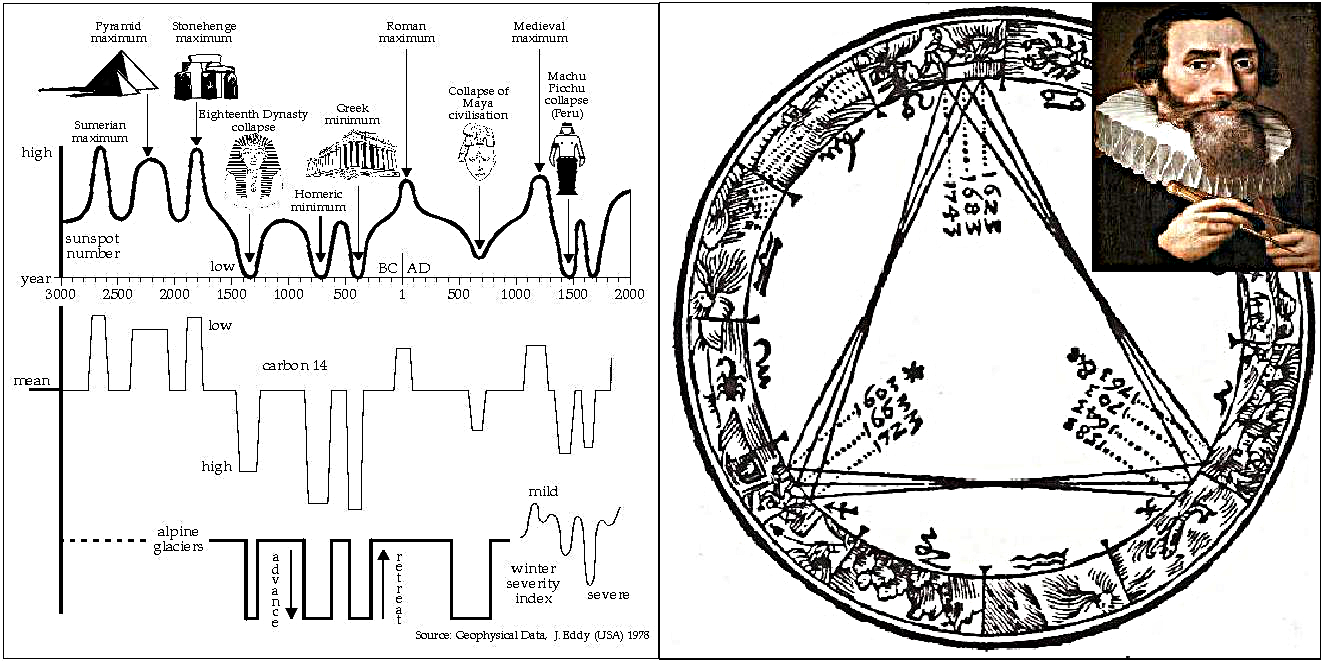}\caption{(Left) Schematic representation of the rise and fall of several civilizations
since Neolithic times that well correlates with the $^{14}C$ radio-nucleotide
records used for estimating solar activity (adapted from Eddy's figures
in Refs. \citep{Eddy,Eddy2}). Correlated solar-climate multisecular
and millennial patterns are recently confirmed \citep{Bond,Kerr,Steinhilber}.
(Right) Kepler's Trigon diagram of the great Jupiter and Saturn conjunctions
between 1583 to 1763 \citep{Kepler1606}, highlighting 20 year and
60 year astronomical cycles, and a slow millennial rotation. }
\end{figure*}

Figure 10 summarizes the performance of one of the CMIP5 GCMs, namely
the CanESM2 GCM, and the typical problems inherent to all CMIP5 GCMs:
(1) GCMs fail to reproduce the 2000-2012 steady GST trend; (2) the
amplitude of the volcanic cooling spikes is too large in the GCMs
compared to their GST signature; (3) the GCMs show for 1880-1960 steady
warming whilst the GST shows a clear 60-year modulation consisting
of a 1880-1910 cooling plus a 1910-1940 warming; (4) after 1950 the
GCMs require a strong aerosol cooling effect to partially compensate
for strong GHG warming up to 2000; since 2000 aerosol cooling no longer
compensates for strong GHG warming to the end that the simulations
strongly diverge from observations.

\section{The ancient understanding of climate change}

For millennia the traditional understanding was that the climate system
is largely regulated by numerous natural oscillations of astronomical
origin working at multiple time scales \citep{Ptolemy,Kepler,Masar,Iyengar}.
For example, soli-lunar calendars were widely used in antiquity because
ancient civilizations considered soli-lunar cycles important for farming
activities: in North America this tradition has been continued since
1792 by the Old Farmer's Almanac.%
\footnote{\href{http://www.almanac.com/}{http://www.almanac.com/}%
} Moreover, quasi 20-year and 60-year astronomical oscillations were
well known too. Ancient civilizations believed that somehow the economy
was related to these astronomical oscillations through the climate
\citep{Ptolemy,Masar,Kepler,Temple,Horrox}. Indeed, cycles with periods
of 7-11 years (Juglar), 15-25 yeas (Kuznets) and 45-60 years (Kondratiev)
have been found among the business cycles.%
\footnote{\href{http://en.wikipedia.org/wiki/Business_cycle}{http://en.wikipedia.org/wiki/Business\_{}cycle}%
} A 60-year cycle was included in Chinese and Indian traditional calendars
probably because these cycles were and are also reflected in the monsoon
cycles \citep{Iyengar,Agnihotri}. In Hindu tradition the 60-year
calendar cycle was referred to as the Brihaspati (= Jupiter) cycle.
In 886 AD \citet{Masar} attempted a comprehensive interpretation
of history based mostly on Jupiter-Saturn oscillations. \citet{Kepler1606},
who strongly promoted astronomical climatology, designed in 1606 his
famous diagram representing these two multidecadal cycles (Fig. 11,
right).

As Kepler's diagram shows, quasi 20 year and 60 year oscillations
could be readily deduced from the orbital period of Jupiter (11.86
year) and Saturn (29.46 year). The Jupiter-Saturn conjunction period
is $\sim19.85$ year, at $\sim$242.57$^{o}$ of angle. Every $\sim$60
years a conjunction Trigon completes with a $\sim$7.7$^{o}$ rotation
(Fig. 7). The full astronomical configuration repeats every $\sim$900-960
years using the sidereal orbital periods of the planets, as \citet{Masar}
observed following Ptolemy, or every $\sim$800 years using the tropical
orbital periods, as \citet{Kepler1606} observed. In both cases, the
slow rotation of the Trigon convinced ancient civilizations of a quasi-millennial
astronomical cycle that could be approximately correlated with a quasi-millennial
cycle commonly observed in historical chronologies, as revealed by
the rise and fall of civilizations. These events were likely driven
by climatic variations (Fig. 11, left) \citep{Eddy,Eddy2}.

Indeed, in 1345 AD a Jupiter-Saturn conjunction occurred in the zodiac
sign of Aquarius and was linked to the outbreak of the Black Death
epidemic \citep[pp. 158-172]{Horrox}%
\footnote{Traditional medieval astrology claimed that when the Trigon of the
great conjunctions of Jupiter and Saturn occurred in the zodiac air
sign of Aquarius \textit{kingdoms have been emptied and the earth
depopulated} because of \textit{great cold, heavy frosts and thick
clouds corrupting the air} \citep[ pp. 172]{Horrox}.%
}. In 1606 \citet{Kepler1606} used a related argument to predict that
European civilization would have flourished again during the following
four/five centuries, and Newton excluded the possibility of another
civilization collapse before 2060 AD%
\footnote{\href{http://en.wikipedia.org/wiki/Isaac_Newton's_occult_studies}{http://en.wikipedia.org/wiki/Isaac\_{}Newton's\_{}occult\_{}studies}%
}. Today, it is known that this quasi-millennial civilization cycle
is also reflected in the $^{14}C$ and $^{10}Be$ cosmo-nucleotide
records, which are modulated by solar activity \citep{Bond,Kerr,Steinhilber}
(Fig. 11) suggesting a planet-sun-climate link.

However, since the 18th century a planetary influence on the climate,
as well as the ancient astrological planetary models were dismissed
as superstitions because according Newton's gravitational law the
planets are too far from the Earth to have any observable effect.
Is there a solution to this curious mystery? Below a modern astronomical
interpretation of the climate oscillations is given based on some
of the author's studies \citep{Scafetta2010,Scafetta2012a,Scafetta2012c,Scafetta2012d,Scafetta20131}.

\section{Planetary control on solar and climate change oscillations throughout
the Holocene}

In the 19th century an important discovery was made: the Sun oscillates
with an apparent 11-year periodicity. In 1859 \citet{Wolf} postulated
that \textit{the variations of the sunspot-frequency depends on the
influences of Venus, Earth, Jupiter and Saturn}. The theory of a planetary
influence on solar activity was popular in the 19th and earlier 20th
century \citep{Brown,DelaRue}. Although most solar physicists no
longer favor this concept claiming that, e.g., planetary forces are
too weak to influence solar activity (for a summary of objections
see Ref. \citep{Scafetta2012c,Scafetta2012d}), \citet{Scafetta2012c,Scafetta2012d}
developed it further using physics and data not yet available in the
19th century, found strong supporting empirical evidence for it and
proposed a physical explanation.

Indeed, a number of recent studies since the 1970s noted correlations
between the wobbling of the Sun around the center of mass of the solar
system and climatic patterns \citep{Charv=00003D0000E1tov=00003D0000E1,Fairbridge,Landscheidt,Abreu,Jakubcov=00003D0000E1}.
However, as the Sun is in free fall with respect to the gravitational
forces of the planets, its wobbling should not effect its activity.
Scafetta \citep{Scafetta2010,Scafetta2012a,Scafetta2012c,Scafetta2012d,Scafetta20131}
investigated this conundrum taking the following four aspects into
consideration:

\begin{figure*}[!t]
\centering{}\includegraphics[width=1\textwidth,height=0.6\paperheight]{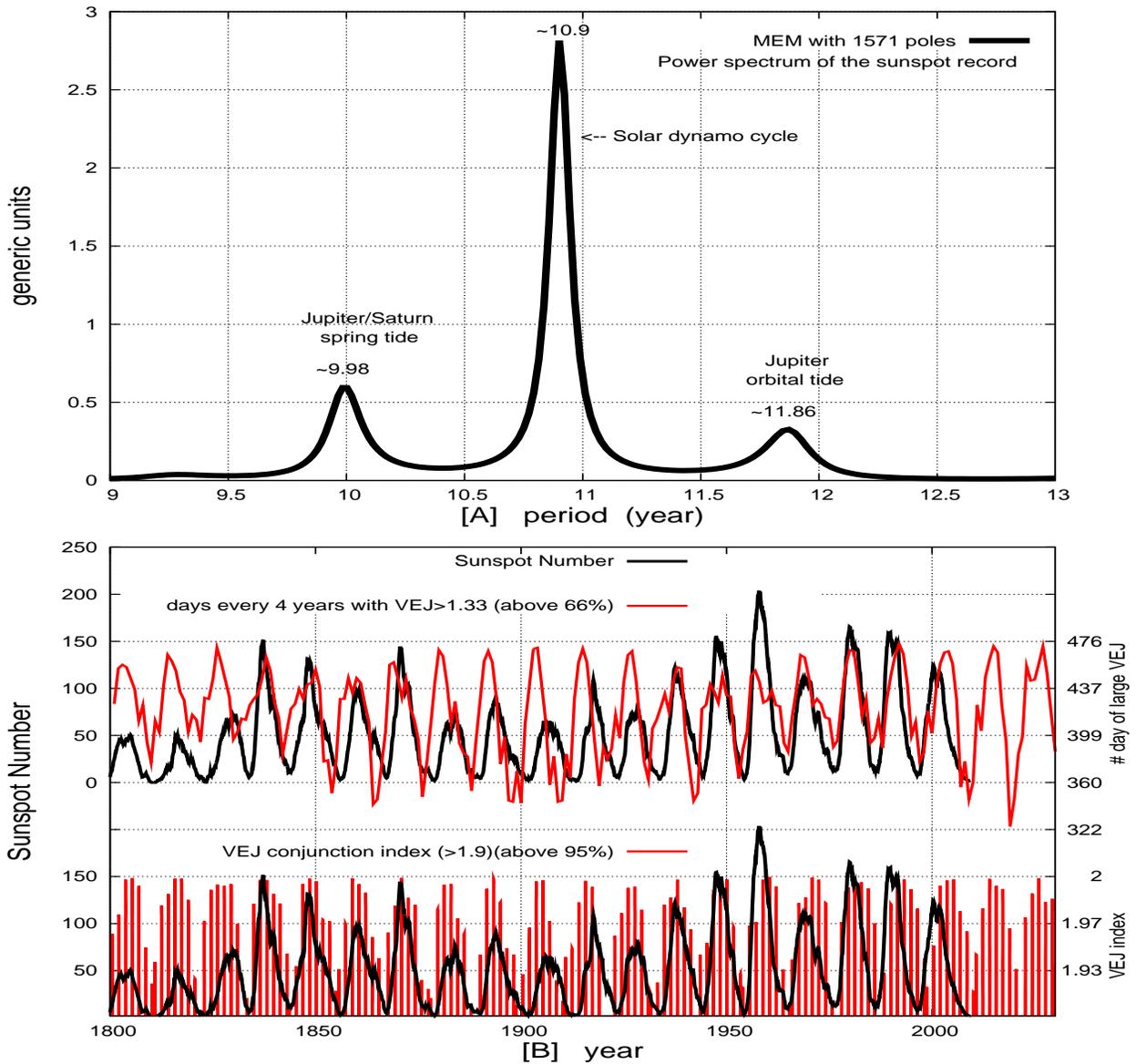}\caption{{[}A{]} Maximum entropy method (MEM) power spectrum of the sunspot
record from 1749 to 2010 highlighting three peaks within the Schwabe
frequency band (period 9-13 years) including the two major tides of
Jupiter and Saturn. {[}B{]} Comparison between the sunspot record
(black) and a particular tidal pattern configuration (red) made using
Venus, Earth and Jupiter (VEJ) that reproduces on average the solar
cycle length of 11.08 year . For details see \citet{Scafetta2012c,Scafetta2012d}.}
\end{figure*}

\begin{enumerate}
\item Using Taylor's theorem \citet{Scafetta2010} explained that even if
the solar wobbling functions are not the direct physical cause of
the observed effects, they can still be used as proxies because they
would present frequencies and geometric patterns in common with the
relevant physical functions even if the latter may remain unknown. 
\item The gravitational and electro-magnetic properties of the heliosphere
may be modulated by the reciprocal position and speed of the Jovian
planets and of the Sun. Moreover, as solar wobbling is real relative
to the Milky Way Galaxy, its velocity may modulate the incoming cosmic
ray flux. Oscillating electro-magnetic properties of the heliosphere,
of the solar wind and of the incoming cosmic ray flux probably cause
climatic oscillations by means of a cloud cover modulation \citep{Kirkby,Svensmark}
and other electro-magnetic mechanisms. 
\item The Jovian planets may periodically perturb the orbit of the Earth,
causing specific climatic oscillations as it happens with the multi-millenial
Milankovic cycles \citep{Roe} (eccentricity, 100,000-year cycle;
axial tilt, 41,000-year cycle; precession, 23,000-year cycle), which
are responsible for intermittent great glaciations. However, decadal-to-millennial
scale orbital perturbations appear to be too small ($\sim1000$ kilometers)
to explain the decadal-to-millennial scale climatic oscillations by
variations in the Earth-Sun distance because the Earth orbits the
Sun and not the barycenter. Nevertheless, this hypothesis needs to
be further investigated. 
\item As discussed in \citet{Scafetta2012c,Scafetta2012d}, a possible physical
mechanism are planetary gravitational tidal forces. Power spectra
of the sunspot record demonstrated that the 11-year Schwabe sunspot
cycle is made up by three interfering cycles, which can be interpreted
as due to: (1) the 9.93-year spring tidal cycle between Jupiter and
Saturn; (2) the 11.86-year Jupiter orbital tidal cycle; (3) a central
oscillation of about 10.87 year that is almost, but not precisely,
the average between the two tidal cycles and may emerge from the solar
dynamo cycle as a collective synchronization harmonic. \citet{Scafetta2012d}
also noted that there are gravitational recurrence patterns of about
11.07-11.08 years due to the Mercury-Venus system and the Venus-Earth-Jupiter
system, which correspond to the average solar cycle length. Thus,
numerous planetary tidal oscillations resonate around the 11-year
Schwabe solar cycle, as postulated by \citet{Wolf}. See Figure 12. 
\end{enumerate}
\begin{figure*}[!t]
\centering{}\includegraphics[width=1\textwidth,height=0.6\paperheight]{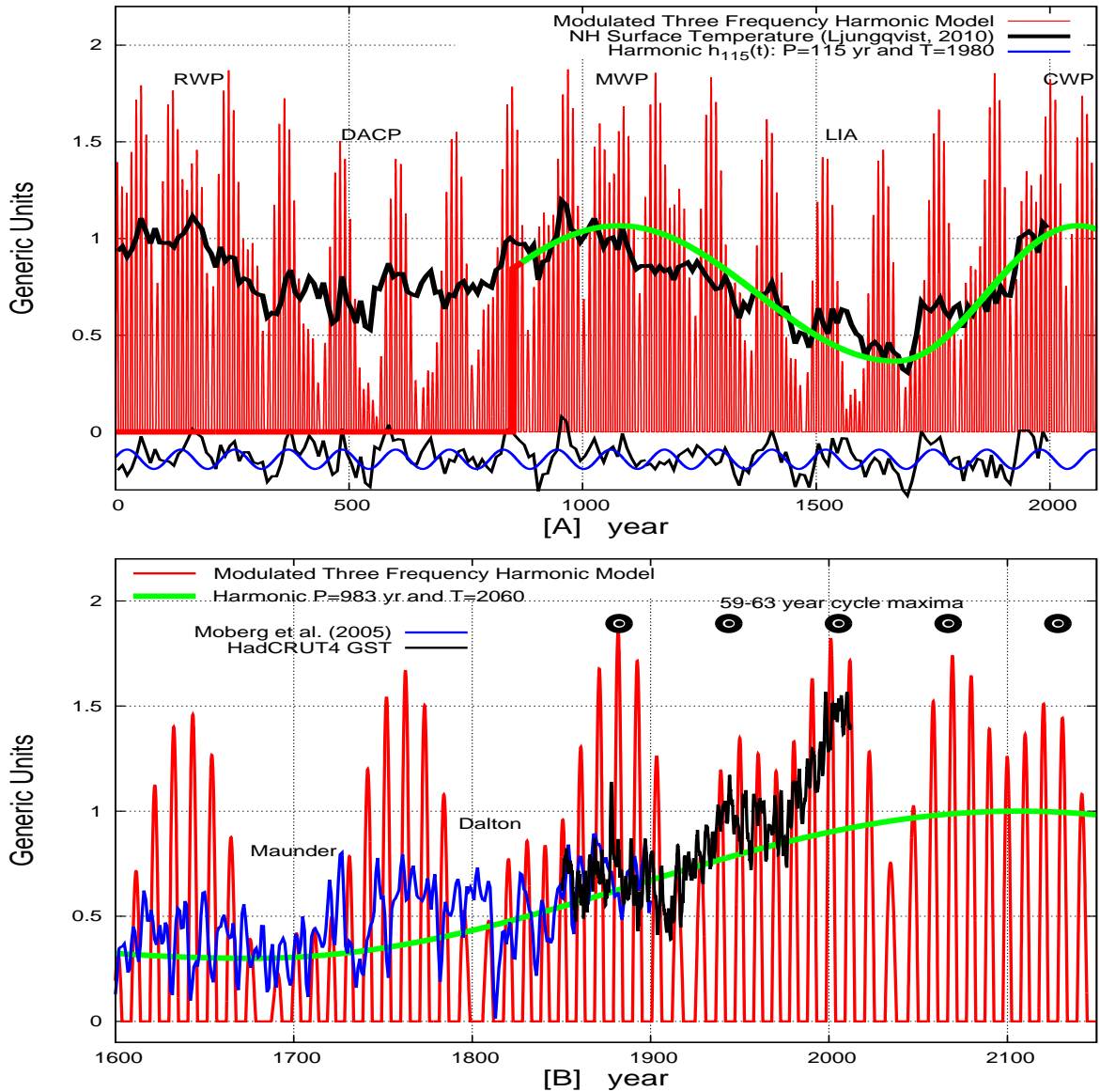}\caption{{[}A{]} Three frequency solar/planetary harmonic model (red) vs. the
Norther Hemisphere temperature reconstruction by \citet{Ljungqvist}
(black). RWP: Roman Warm period. DACP: Dark Ages Cold Period. MWP:
Medieval Warm Period. LIA: Little Ice Age. CWP: Current Warm Period.
{[}B{]} Same solar model (red) vs. HadCRUT4 GST (annual smooth: black)
combined in 1850-1900 with the proxy temperature model of \citet{Moberg}
(blue). Green curve: millennial modulation (Eq. \ref{eq:6}). After
\citet{Scafetta2012c}.}
\end{figure*}

Taking these considerations into account, a simple solar model was
developed by \citet{Scafetta2012c} involving just three harmonics,
namely the two Jupiter/Saturn tidal cycles and a hypothetical solar
dynamo cycle with a 10.87-year period. This model reproduces a varying
11-year cycle that correlates approximately with the Schwabe sunspot
cycle, and produces beats at about 61 years, 115 years, 130 years
and 983 years, which are synchronous with major solar and climatic
multidecadal, secular and millennial oscillations observed throughout
the Holocene. The model recovers: (1) the quasi millennial oscillation
observed in both solar and climate proxy records \citep{Bond,Kerr};
(2) the prolonged periods of low solar activity during the last millennium
know as the Oort, Wolf, Spörer, Maunder and Dalton minima; (3) the
seventeen 115 year oscillations found in detailed temperature reconstructions
of the Northern Hemisphere covering the last 2000 years \citep{Ljungqvist,Qian}
that correlate with periods of grand solar minima (see bottom of Figure
13A); (4) the quasi 61-year GST modulation that has been clearly observed
in GST records since 1850.

Figure 13 \citep{Scafetta2012c} shows that a modulated quasi 61-year
beat oscillation apparently dominates solar dynamics between 1850
and 2150 AD. Other solar harmonics such as the $\sim$87-year Gleissberg
and the $\sim$207-year de Vries solar cycles and other solar cycles
can be readily discerned in the planetary harmonics \citep{Scafetta2012d,Scafetta20131,Abreu}.
In fact, the 1/7 resonance of Jupiter and Uranus is about 85 years
other harmonics occur at 84-89 years \citep{Jakubcov=00003D0000E1,Scafetta20131};
the beat resonance between the quasi 60-year and the 85-year cycles
is about 205 years. Solar variation is likely the result of an internal
complex collective synchronization dynamics \citep{Pikovsky} emerging
from numerous gravitational and electro-magnetic harmonic forcings.

\citet{Scafetta2012c} estimated that the 115-year cycle should peak
in 1980 and the 983-year cycle in 2060. By using the paleoclimate
temperature records given in Figures 3 and 11A \citep{Ljungqvist,Moberg},
and by looking at the cooling between the Medieval Warm Period (MWP)
ending around 1000 AD and the Little Ice Age (LIA) around 1670 AD,
it can be deduced that the amplitude of the 115-year cycle is about
$0.1\pm0.05$ $^{o}C$ and that the millennial cycle amplitude is
about $0.7\pm0.3$ $^{o}C$. The millennial climatic cycle appears
to have reached also a minimum around 1680: see also \citet{Humlum}.
Therefore, the two cycles can be approximately represented as:

\begin{equation}
h_{115}(t)=0.05\cdot\cos(2\pi(t-1980)/115)\label{eq:5}
\end{equation}

\begin{equation}
h_{983}(t)=0.35\cdot\cos(2\pi(t-2060)/760)\label{eq:6}
\end{equation}
Eq. \ref{eq:6} uses the period of 760-year for simulating the skewness
of the millennial climatic cycle and should be valid from 1680 to
2060.

\section{Sun as amplifier of planetary orbital oscillations }

\citet{Scafetta2012d} proposed the following physical mechanism to
explain how planetary forces may modulate solar activity. Planetary
tides on the Sun are extremely small, and therefore scientists were
discouraged from believing that these regulate solar activity. However,
systems that generate energy can work as amplifiers and, evidently,
the Sun is a powerful generator of energy. Since its nuclear fusion
activity is regulated by gravity, the Sun may work as a huge amplifier
of the small planetary gravitational tidal perturbations exerted on
it.

Solar luminosity is related to solar gravity via the well-known Mass-Luminosity
relation: if the mass of the Sun increases, its internal gravity increases
and makes more work on its interior masses. Consequently, solar luminosity
increases as: $L/L_{S}\approx(M/M_{S})^{4}$ \citep{Duric}. For example,
if the mass of all planets were added to the Sun, the total solar
irradiance would increase by about 8 $W/m^{2}$. Using an argument
based on the Mass-Luminosity relation, \citet{Scafetta2012d} estimated
that nuclear fusion could greatly amplify the weak gravitational tidal
energetic signal dissipated inside the Sun by up to a 4 million factor.
If this is so, planetary tides are able to trigger solar luminosity
oscillations with a magnitude compatible with the observed TSI oscillations
\citep{ScafettaWillson2009}, and, consequently, may be able to modulate
the solar dynamo cycle. Alternative planetary mechanisms influencing
the Sun may exist; see also \citet{Wolff2010} and \citet{Abreu}.

\section{Total Solar Irradiance (TSI) uncertainty problem}

The three-frequency solar model (Fig. 13) predicts, akin to the GST,
that solar activity increased from 1970 to 2000, reached a maximum
around 2000, and will decrease until the 2030s. However, the CMIP5
GCMs adopt a solar forcing deduced from Lean's solar proxy models
\citep{Wang,Kopp} that show a flat TSI trend since 1955 with a slight
decrease since 1980. Also the CMIP3 GCMs used by the \citet{IPCC}
adopted Lean's TSI models in an effort to show that during the last
40 years a more or less stable Sun could not be responsible for the
observed warming after the1970s: see also \citet{Lockwood}. This
is a highly controversial issue that needs to be clarified.

\begin{figure*}[!t]
\centering{}\includegraphics[width=1\textwidth,height=0.6\paperheight]{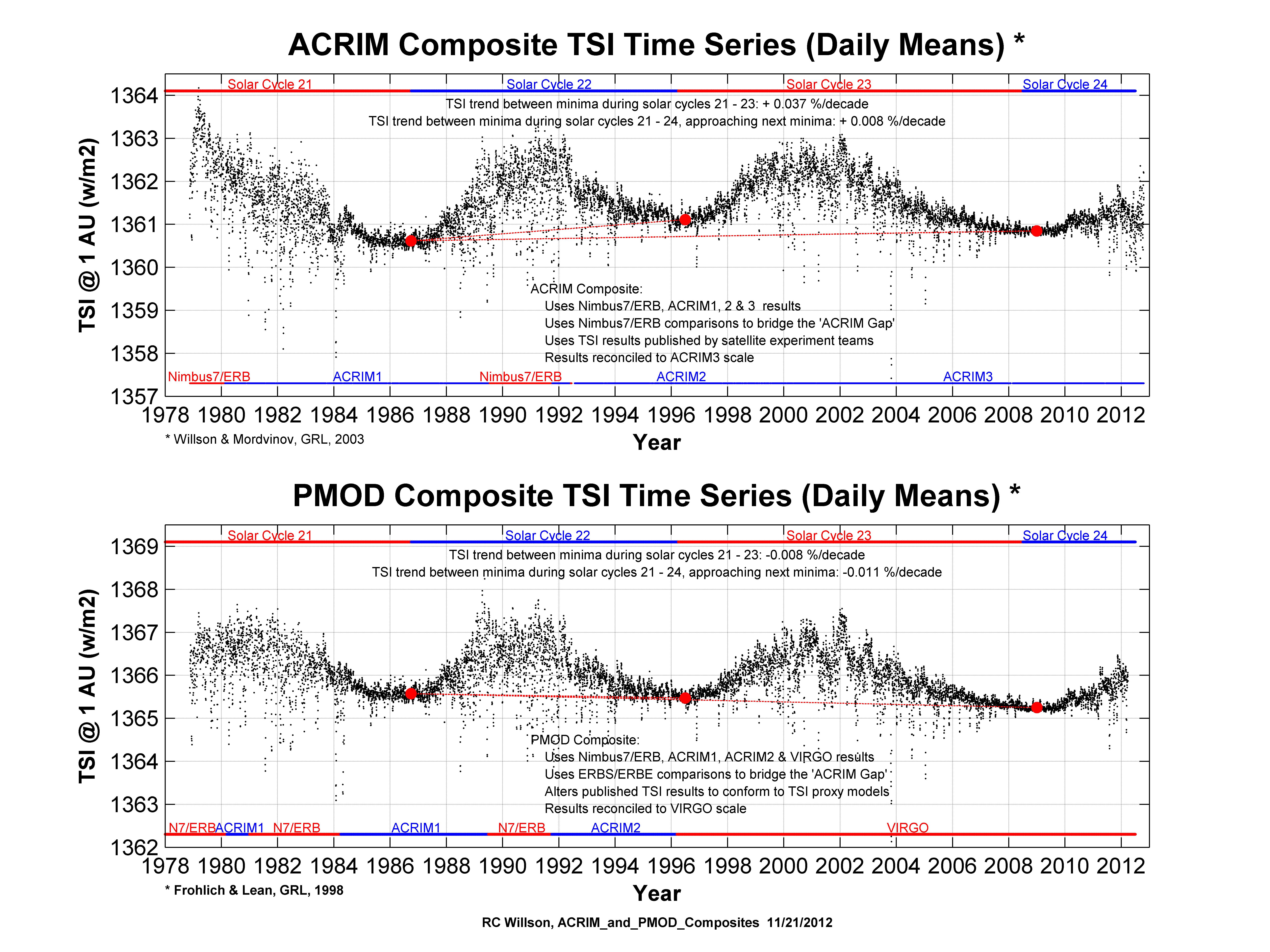}\caption{(Top) ACRIM total solar irradiance satellite composite \citep{Willson2003}.
(Bottom) PMOD total solar irradiance composite \citep{Frohlich}. }
\end{figure*}

It is claimed that Lean's proxy models are supported by actual TSI
observations provided by the PMOD satellite TSI composite \citep{Frohlich1998,Frohlich}.
However, PMOD used \textit{modified} TSI satellite records. For some
unexplained reason, the scientific community appears to ignore that
not only have the experimental teams responsible for the TSI satellite
data never validated these PMOD modifications of their records, but
explicitly stated that PMOD's procedures are highly speculative and
physically incompatible with the experimental recording equipments
\citep{ScafettaWillson2009,Scafetta2011} (see the Appendix). By contrast,
the ACRIM TSI satellite composite \citep{Willson2003} is based on
TSI satellite data as published, and shows that TSI increased between
1980 and 2000 and decreased thereafter. Figure 14 compares the ACRIM
and PMOD TSI composites. With simple empirical thermal models \citet{Scafetta2007}
and \citet{Scafetta2009,Scafetta2011} assessed the implication of
adopting the ACRIM and PMOD TSI composites and showed that with the
ACRIM record most of the climate change observed since the Maunder
Minimum, including the 1970-2000 warming, can be attributed to variations
in solar activity.

\begin{figure*}[!t]
\centering{}\includegraphics[width=1\textwidth,height=0.6\paperheight]{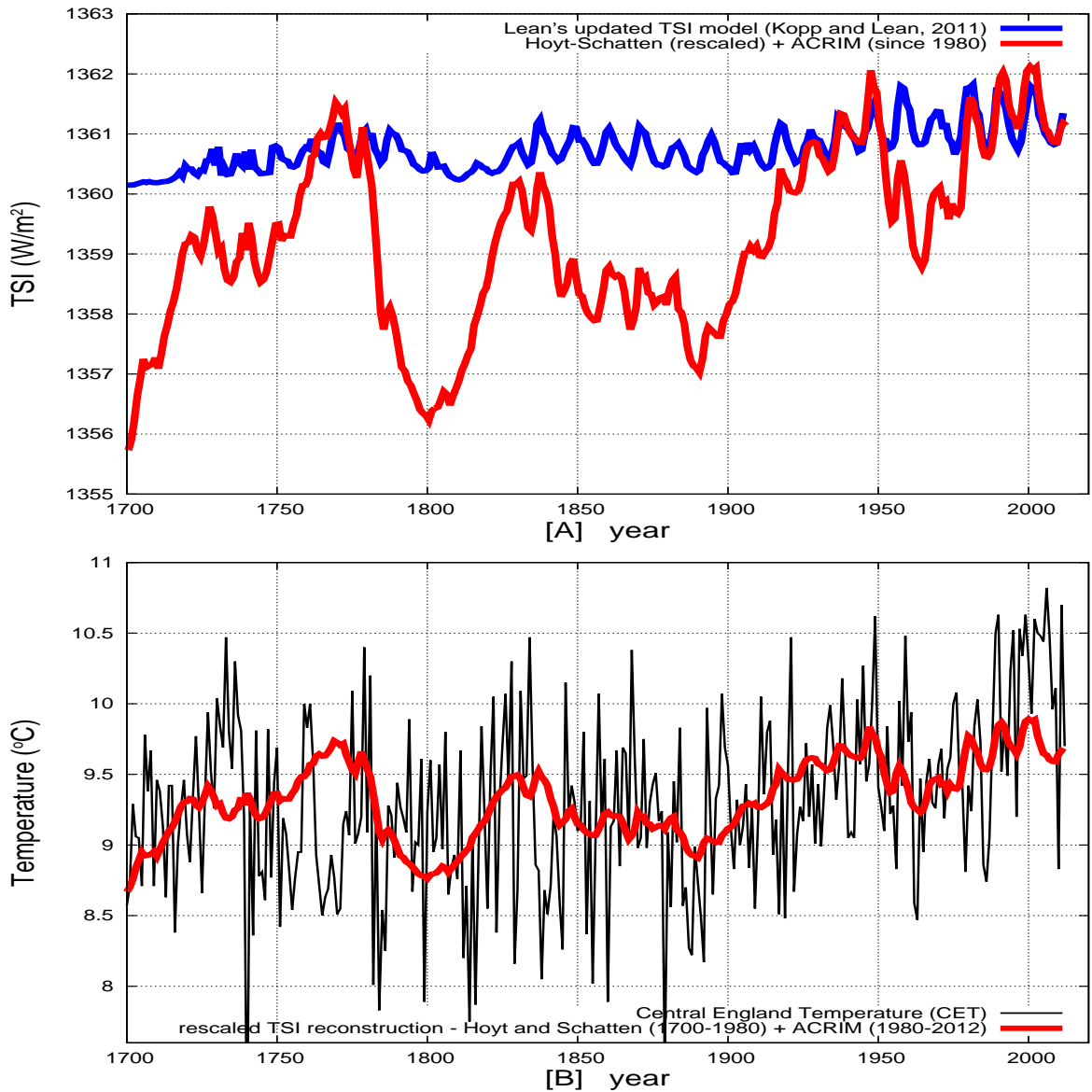}\caption{{[}A{]} Total solar irradiance (TSI) reconstruction by \citet{Hoyt,Hoyt1993}
updated with the ACRIM record \citep{Willson2003} (since 1980) (red)
vs. the updated Lean's model \citep{Wang,Kopp} (blue) used as solar
forcing function in the CMIP5 GCMs adopted in the IPCC AR5 in 2013.
{[}B{]} Comparison between the Central England Temperature (black)
\citet{Parker} and the TSI model by \citet{Hoyt1993} plus the ACRIM
TSI record.}
\end{figure*}

The controversy between the ACRIM and PMOD composites centers mainly
on the TSI trend during the so-called \textit{ACRIM-gap} of 1989-1992.
PMOD claims that during this period the Nimbus7/ERB TSI record, which
is necessary to bridge the ACRIM1 and ACRIM2 TSI records, must be
shifted and inclined downward to produce by 1992 a total downward
shift of about 0.8-0.9 $W/m^{2}$ \citep{Frohlich,Scafetta2011}.
This modification of the Nimbus7/ERB TSI record results in a decreasing
TSI trend during 1989 to 1992 and causes in the PMOD TSI composite
the 1996 TSI minimum to be almost at the same level as the 1986 TSI
minimum \citep{Scafetta2011}. On the contrary, the ACRIM way of combining
the TSI records without modifying them implies that the 1996 minimum
was about 0.5 $W/m^{2}$ higher than the 1986 TSI minimum: this difference
would be about 0.8-0.9 $W/m^{2}$ by adopting the same composite PMOD
merging methodology \citep{Scafetta2011}. Thus, without PMOD's modification,
TSI clearly increased during 1980-2000, similar as the GST \citep{Scafetta2007,Scafetta2009}.

Most arguments supporting the PMOD TSI composite are based on highly
controversial proxy models such as Lean's models and a few others
\citep[e.g.: ][]{Wang,Frohlich1998,Krivova2007,Ball}, which agree
with the flat PMOD TSI pattern partially employing circular reasoning.
However, a correct scientific argument must focus on the ACRIM-gap
data to which the most important PMOD modifications are applied. Once
this is done \citet{ScafettaWillson2009} showed that the model of
\citet{Krivova2007} is not compatible with the PMOD modification
of Nimbus7/ERB record. \citet{Krivova2009} did not query this, but
remarked that other proxy models, claimed to be more accurate on short
time scales, confirm the PMOD TSI composite. However, a direct comparison
of the smoothed ACRIM and PMOD TSI composites and of the magnetogram-based
SATIRE TSI model during the ACRIM-gap, as given by \citet[fig. 8]{Ball},
shows that, similar to ACRIM, from 1990 to 1992.5 also SATIRE trends
upward from 1990 to 1992.5 (slope = $0.1\pm0.03$ $Wm^{-2}/year$)
approximately as in the original Nimbus7/ERB measurements. Also the
Climax Neutron Monitor cosmic ray intensity count, which inversely
correlates with solar magnetic activity, decreases between 1989 and
1992 \citep{Scafetta2011}. In general, the average cosmic ray flux
steadily decreased between 1965 and 1996 \citep{Kirkby}. Therefore,
preference should be given to the ACRIM TSI composite that is probably
closer to reality than the PMOD TSI composite.

In addition, while the three-frequency solar model presents a maximum
in the 1940s that clearly correlates with a temperature maximum (Fig.
13), Lean's solar proxy models \citep{Frohlich1998,Wang} peak in
1960 similar to the sunspot number record. However, the solar maximum
of the 1940s is supported by the TSI reconstruction of \citet{Hoyt}
and by the solar cycle length model \citep{Loehle2011,Thejll}. Indeed,
since 1900 the TSI reconstruction of \citet{Hoyt}, updated by Scafetta
with the ACRIM record, clearly correlates with the 60-year modulation
of the GST records during the 20th century. Soon \citep{Soon2005,Soon,Soon2013}
used the updated \citet{Hoyt} TSI model to demonstrate its good correlation
with the GSTs of the Arctic and China, with the Japanese sunshine
duration record and with the Equator-to-Pole (Arctic) temperature
gradient function. Figure 15A compares these two alternate TSI models
\citep{Hoyt,Kopp} and highlights the severity of their difference
particularly in terms of TSI multidecadal variation. Figure 15B compares
the Central England Temperature (CET) record \citep{Parker} and the
TSI model by \citet{Hoyt1993} plus the ACRIM TSI record; an overall
good correlation is observed since 1700, which suggests that the major
observed climatic oscillations are solar induced and that the Sun
explains about 50-60\% of the warming observed since 1900.

The good multisecular correlation between CET and the chosen secular
TSI reconstruction, which includes the quasi 60-year climatic oscillation
observed since 1900, contradicts a claim of \citet{Zhou} and \citet{Tung2013}.
These authors have inappropriately criticized \citet{Scafetta2005,Scafetta2006}
claiming that solar activity has contributed less than 10\% of the
warming for the first half of the 20th century. In their opinion the
observed warming was mostly induced by the Atlantic Multidecadal Oscillation
(AMO) during its 1910-1940 warm phase. However, the AMO is not an
independent forcing of the climate system but it is a subsystem of
the global temperature network and of the climate system itself. The
result depicted in Figure 15B clearly suggests that the observed climatic
oscillations, including those of its subsystems such as AMO, are driven
by solar activity \citep{Scafetta2010,Scafetta2012a,Mazzarella,Loehle2011,Soon2005,Soon2013,Hoyt1993}.

The TSI model proposed by \citet{Hoyt,Hoyt1993} uses various indexes
such as sunspot cycle amplitude, sunspot cycle length, solar equatorial
rotation rate, fraction of penumbral spots, and the decay rate of
the approximate 11-year sunspot cycle. On the contrary, Lean's models
are based mostly on regression of sunspot blocking and faculae brightening
indexes. The latter indexes may not well capture the dynamics of the
quiet-sun background radiation that may be responsible for a multidecadal
TSI trending, as observed in the ACRIM composite. Similarly, \citet{Shapiro}
and \citet{Judge} recently argued that multi-decadal TSI variations
are significantly larger than suggested by Lean's models because the
quiet-sun magnetic field may vary significantly on multidecadal time
scales following the modulation potential of the galactic cosmic rays.

In conclusion, it is possible that current GCMs are using an erroneous
solar input function owing to their adoption of the Lean's TSI models
\citep{IPCC,Wang,Kopp}, which may be flawed and, therefore, may severely
obscure the true solar effect on climate.

\section{Empirical astronomical model}

The six climatic harmonics synchronous with astronomic cycles at decadal
to the millennial time scales are approximated by Eqs. 1-6. These
harmonics describe phenomenologically the natural climatic oscillations
observed since 1850 and, as demonstrated above and in Ref. \citep{Scafetta2012b},
are not reproduced by the GCMs. These functions permit empirical modeling
of natural variability at decadal to millennial time scales. However,
GST also depends on the chemical composition of the atmosphere that
can be modified by anthropogenic emissions and volcano activity.

As discussed in the introduction, the \citet{IPCC} concluded that
100\% of the $\sim$0.5-0.55 $C^{o}$ warming observed since 1970
can only be explained by anthropogenic forcing. However, its adopted
GCMs fail to reproduce the natural harmonics such as the 60-year oscillation
(Fig. 5). \citet{Scafetta2012b} argued that the failure of the GCMs
to properly reproduce the 60-year oscillation, which contributed about
0.3 $^{o}C$ of warming between 1970 and 2000, caused the GCMs to
overestimate the climatic effect of anthropogenic forcing by about
50-60\%. \citet{Zhou} reached a similar conclusion by using the Atlantic
Multidecadal Oscillation that shows a clear 60-year oscillation \citep[fig. 9B]{manzi}
without realizing that climatic oscillations such as AMO may be astronomical/solar
induced. Even assuming that the anthropogenic forcing functions used
in the GCMs are correct, the above arguments imply that the GCMs significantly
overestimated the climate sensitivity to radiative forcing because
their output ought to be reduced to about a factor of 0.45. Such a
reduction is in keeping with modern paleoclimatic reconstructions
(Fig. 4) and the calculations in section 3. The proposed correction
provides a first approximation estimate of the climatic effect of
the anthropogenic plus volcano forcings, given that according the
CMIP5 GCMs, the solar contribution to the secular GST trend is nearly
insignificant (at best a few percent).

\begin{figure*}[!t]
\centering{}\includegraphics[width=1\textwidth,height=0.6\paperheight]{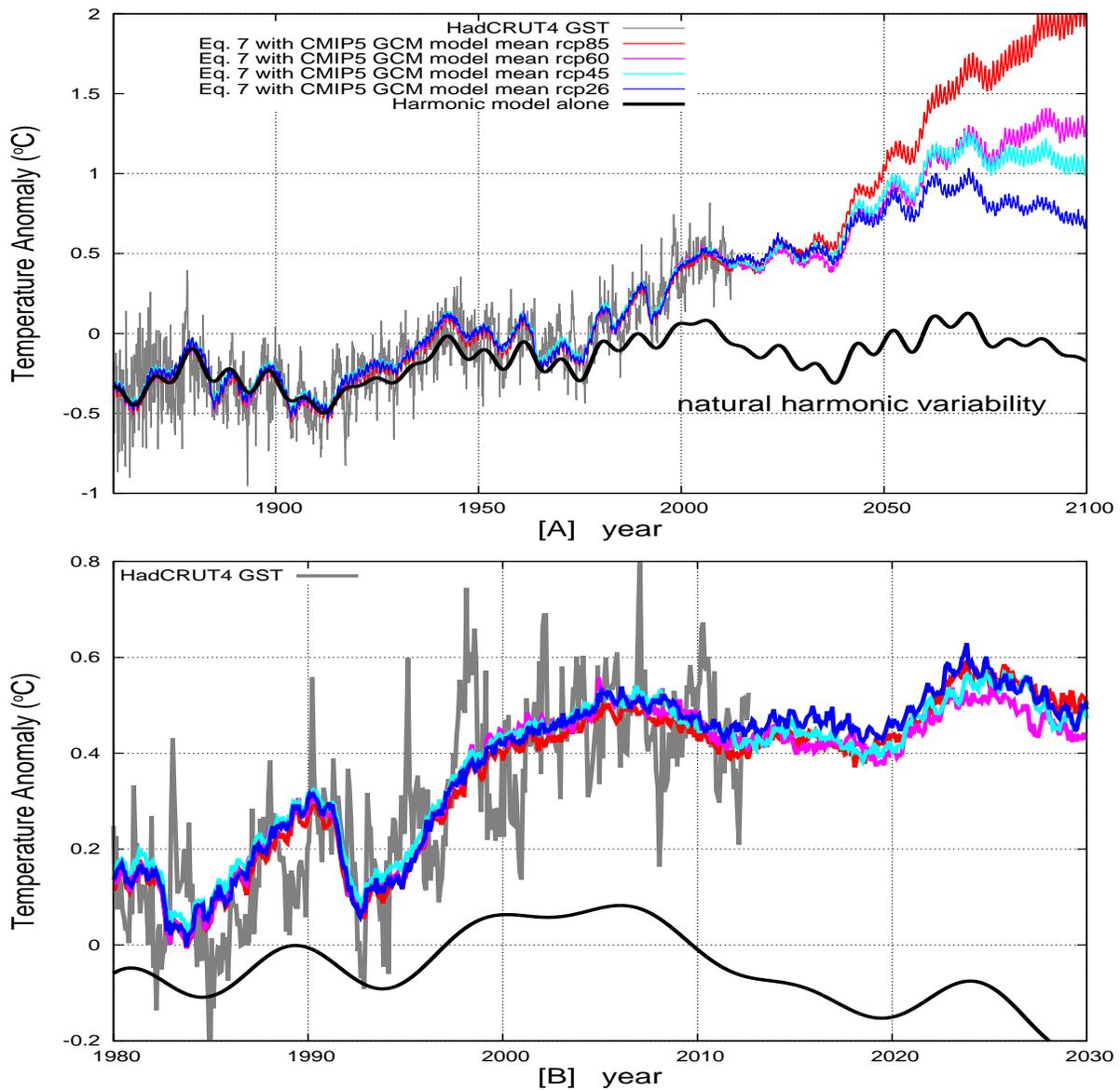}\caption{{[}A{]} HadCRUT4 GST (gray) superimposed on the empirical climate
model given by Eq. \ref{eq:7} where the anthropogenic/volcano component
$M_{A,V}(t)$ is derived from the four alternate CMIP5 GCM ensemble
average simulations of Fig. 1. The smooth black curve corresponds
to the six-frequency harmonic component alone, representing the modeled
natural variability. {[}B{]} Zoom of {[}A{]} for 1980 to 2030. }
\end{figure*}

GST can be phenomenologically modeled with the following equation:

\begin{equation}
\begin{array}{cc}
f(t)= & h_{9.1}(t)+h_{10.2}(t)+h_{21}(t)+h_{61}(t)\\
 & +h_{115}(t)+h_{983}(t)+M_{A,V}(t),
\end{array}\label{eq:7}
\end{equation}
where the natural harmonic component of the climate system is reconstructed
by the six harmonics that are presumably induced by synchronized astronomical
harmonic forcings. The function $M_{A,V}(t)=0.45\cdot m_{GCM}(t)$
is the output of the GCM ensemble averages, $m_{GCM}(t)$, given in
Figure 1, reduced to a 0.45 factor. In first approximation $M_{A,V}(t)$
would simulate the anthropogenic and the short-scale strong volcano
effects on climate under the assumption that the true equilibrium
climate sensitivity to radiative forcing is 55\% smaller than the
one currently simulated by the GCMs.

The empirical climate model of Eq. \ref{eq:7} may require additional
harmonics to include the effects of other minor oscillations and some
additional nonlinear effect. In particular, Figure 13 indicates that
the 61-year solar oscillation appears to be strong in 1850-2150, but
too faint before 1850, while the 115-year oscillation appears to be
stronger before 1850, giving rise to the cold Maunder and Dalton periods.
For simplicity, these corrections are ignored.

Figure 16 gives the HadCRUT4 GST (gray) and the empirical climate
model of Eq. \ref{eq:7} derived from the four alternate CMIP5 GCM
ensemble average simulations and their 21st century projections as
shown in Figure 1. The black curve corresponds to the harmonic model
alone, which represents the natural harmonic variability. The multicolored
curves show $f(t)$ that adequately reproduces both the GST upward
trend since 1850 and all decadal and multidecadal oscillations observed
in the temperature record.

It is important to stress that: (1) the used harmonics are also found
in good synchrony with astronomical cycles; (2) the weight of the
anthropogenic component, that is the factor 0.45 used to reduce the
outputs of the GCM ensemble means, was determined using the period
1970-2000 \citep{Scafetta2012b}; (3) the amplitude of the quasi millennial
cycle was deduced using the cooling from the MWP to the LIA in 1700
AD. Interesting, the correct reconstruction of the warming trending
since 1850 is actually a hind-cast. \citet{Scafetta2012a,Scafetta2012b}
also showed that the harmonic model can be calibrated during the period
1850-1950 to hind-cast the climatic oscillations observed from 1950
to 2010, and vice versa. This demonstrates the hind-cast capability
of the proposed harmonic model and points towards the reliability
of its forecasting capability.

The proposed empirical model hind-casts the GST standstill observed
since 2000, which the GCMs failed to predict, and forecasts a more
or less steady climate oscillating with decadal and bidacadal oscillations
up to 2030-2040 in response to the cooling phase of the 61 year solar/astronomical
cycle that compensates for the projected anthropogenic warming component.
Furthermore, the empirical model predicts a possible warming of 0.3-1.6
$^{o}C$ by 2100 relative to 2000 that is significantly lower than
the 1.1-4.1 $^{o}C$ warming of the GCM ensemble mean projections
given in Figure 1 \citep{Knutti2012}.

\section{Conclusion}

It may be surprising to many to learn that planetary oscillations
probably exert a significant control on the Earth's climate system,
as presented in this paper. However, this is the way climate change
has been interpreted and predicted for millennia by ancient civilizations
that built sophisticated astronomic observatories to this purpose.
\citet{Ptolemy} stated that the motions of the \textit{aether} (that
is the oscillations of the heliosphere driven by the Sun, the Moon
and the planets) alter the uppermost part of the Earth's atmosphere
(which was believed to be made of \textit{fire}), which then alters
the lower atmosphere \textit{acting on earth and water, on plants
and animals} and, consequently, humans are also influenced by the
\textit{stars}. \citet{Kepler,Kepler1606} observed that the climate
had to \textit{respond to the dictates of heavenly harmonies}, and
said that\textit{ nature is affected by an aspect just as a farmer
is moved by music to dance} \citep{Kemp}. Planetary harmonics were
extensively used to interpret human history \citep{Masar} and forecast
monsoon rainfall cycles \citep{Iyengar}. Today, many dismiss this
ancient science as \textit{astrology,} perhaps without realizing that
also calendars and ocean tidal models originated as astrological models,
while today these models are scientifically very well founded. Indeed,
ancient astrology was a mixing of factual facts describing astronomical-geophysical
phenomena and superstitions and, by understanding this, Kepler warned
to not \textit{throw out the baby with the bathwater} \citep{Kepler,Rabin}.

The GST clearly oscillates and increased since 1850. However, the
GCMs used by the IPCC, such as the CMIP3 in 2007 and the CMIP5 in
2013, are unable to reconstruct the observed GST decadal and multidecadal
oscillations. The traditional justification for this failure has been
attributed to an internal variability of the climate system that would
be impossible to properly model due to uncertainties in the initial
conditions and to the chaotic dynamics of the climate system itself.

\begin{figure*}[!t]
\centering{}\includegraphics[width=1\textwidth]{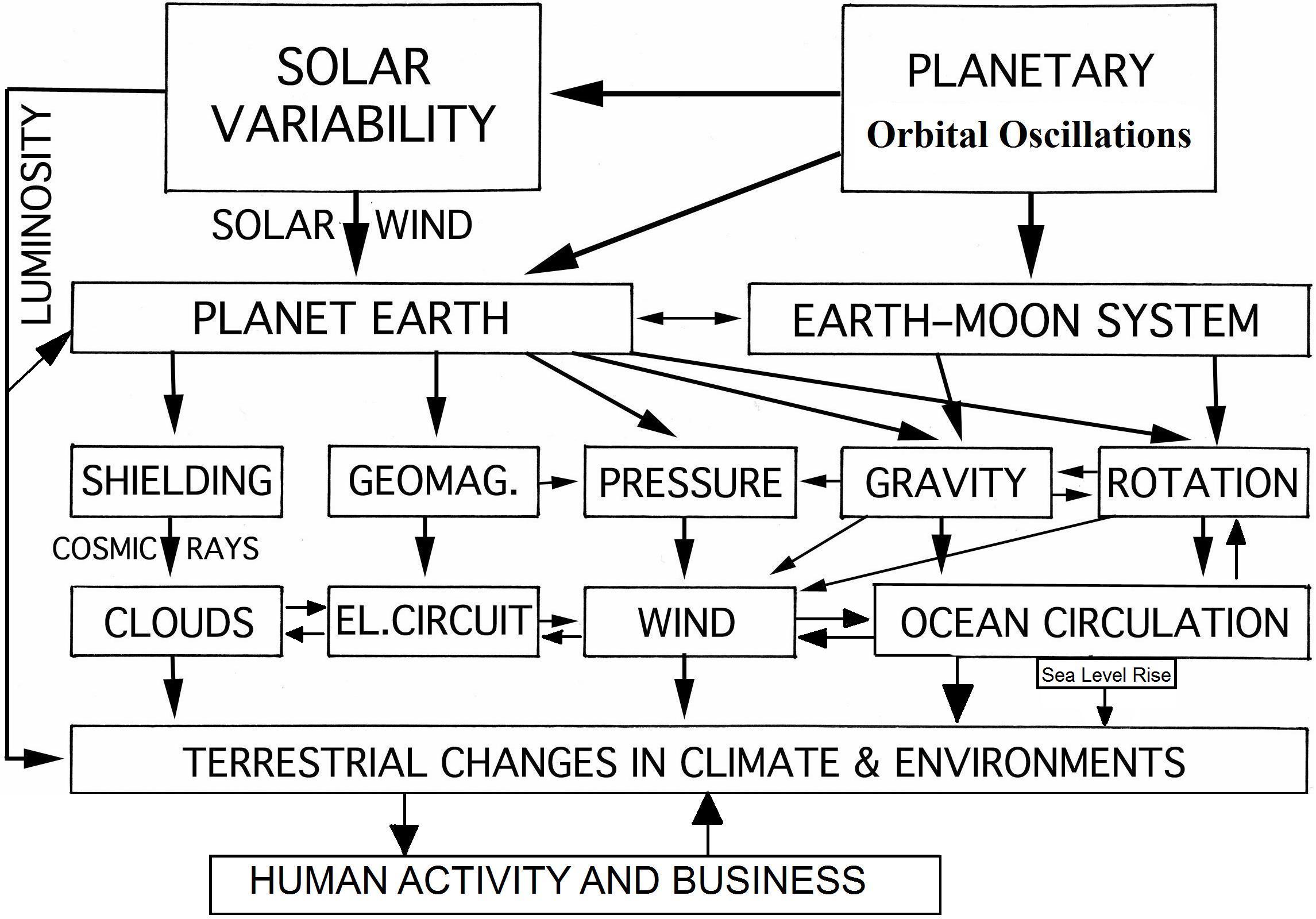}\caption{Network of the possible physical interaction between planetary harmonics,
solar variability and climate and environments changes on Planet Earth
(with permission adapted after \citet{Morner}).}
\end{figure*}

The author \citep{Scafetta2010,Scafetta2012a,Scafetta2012b,Scafetta2012c,Scafetta2012d}
noted that the GST records are characterized by specific frequency
peaks corresponding to astronomical harmonics linked to soli-lunar
tidal cycles, solar cycles and heliosphere oscillations in response
to movements of the planets, particularly of Jupiter and Saturn. Moreover,
he proposed a physical model that may explain how planetary tidal
harmonics can modulate solar activity, and reconstructed the major
known Holocene solar variations \citep{Scafetta2012c,Scafetta2012d}
from the decadal to the millennial scales. Figures 6, 9 and 13 show
that the observed GST and astronomic oscillations are well synchronized.
Indeed, a planetary hypothesis of solar variation is reviving \citep{Charbonneau}.

Empiric harmonic models based on these oscillations are able to reconstruct
all observed major decadal and multidecadal climate variations with
a far greater accuracy than any IPCC AGWT GCMs. A simple harmonic
model based on a minimum of four astronomic oscillations with periods
of about 9.1, 10-12, 19-22 and 59-63 years can readily reconstruct
and hind-cast all so-called GST \textit{hiatus} periods observed since
1850. This contradicts \citet{Meehl} that the observed GST oscillations
are due to an \textit{unpredictable} internal variability of the climate
system.

The full GST record can be reconstructed by using two additional secular
($\sim$115 year) and millennial ($\sim$983 year) astronomical harmonics
plus a climate component regulated by the chemical properties of the
atmosphere (e.g. GHG and aerosols). The GCM ensemble means can be
used to estimate the effect of this component provided that their
output is reduced to a 0.45 factor. Thus, while the current GCMs produce
an average climate sensitivity to $CO_{2}$ doubling to be around
3 $^{o}C$, the real average value of the climate sensitivity may
be approximately 1.35 $^{o}C$ and may very likely vary from 0.9 $^{o}C$
to 2.0 $^{o}C$. The climate sensitivity may be lower, though, provided
part of the GST warming of non-climatic origin, such as uncorrected
urban heat island (UHI) effects \citep{McKitrick2007,McKitrick2010}
or if the preindustrial GST millennial variability is found larger
than what used in Eq. \ref{eq:7} \citep{Christiansen}. This estimate
of low equilibrium climate sensitivity to radiative forcing is compatible
with the results of other studies \citep{Lindzen2011,Spencer2011}.

The empirical model given in Figure 16 implies that about 50-60\%
of the about 0.8-0.85 $^{o}C$ warming observed since 1850 is due
to a combination of natural oscillations, including a quasi-millennial
cycle that was in its warming phase since 1700 AD. Since 1850 major
quasi 20-year and 61-year cycles describe the GST multidecadal scales,
and two decadal cycles of about 9.1 years and 10-11 years capture
the decadal GST scale. Other minor oscillations of about 12, 15 and
30 year period, also linked to astronomical oscillations (see Fig.
6) appear to be present but are ignored here. The proposed six-frequency
empirical climate model, Eq. \ref{eq:7}, outperforms all CMIP3 and
CMIP5 GCM runs, and predicts under the same emission scenarios a significantly
lower warming for the 21st century ranging from 0.3 $^{o}C$ to 1.6
$^{o}C$, which is an upper limit.

\begin{figure*}[!t]
\centering{}\includegraphics[width=1\textwidth,height=0.5\paperheight]{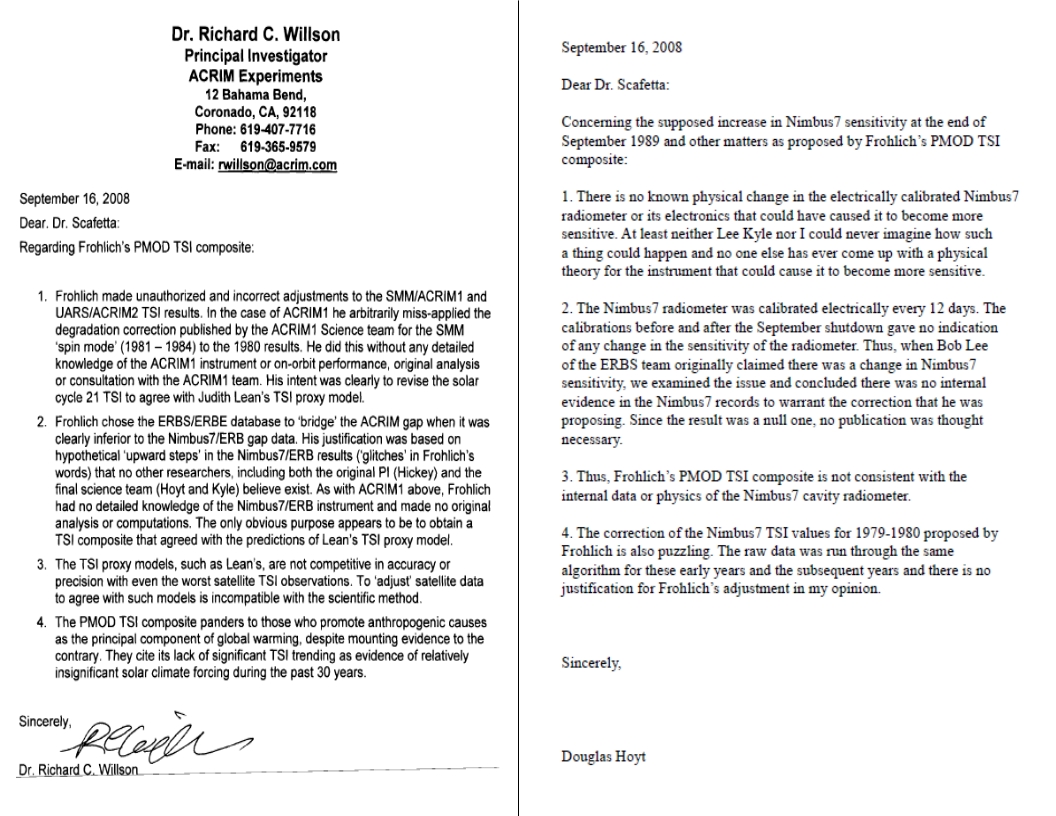}
\caption{Willson and Hoyt's statements regarding the modifications implemented
by Fröhlich \citep{Frohlich1998,Frohlich} to the ACRIM and Nimbus
7 published records. }
\end{figure*}

The results of this analysis indicate that the GCMs do not yet include
important physical mechanisms associated with natural oscillations
of the climate system. Therefore, interpretations and predictions
of climate change based on the current GCMs, including the CMIP5 GCMs
to be used in the IPCC AR5, is questionable. Mechanisms missing in
the GCMs are probably linked to natural solar/astronomical oscillations
of the solar system that are a subject of further research. However,
these oscillations can be already empirically modeled and, in first
approximation, used for forecasting at least the harmonic component
of the climate system.

Figure 17 presents a qualitative diagram summarizing the network of
the possible physical interaction between planetary harmonics, solar
variability, soli-lunar tidal forcings, climate and environmental
changes on our Earth. Taking into account a planetary oscillation
control of solar activity and lunar harmonics controlling direct or
indirect natural climatic forcings, may make solar and climate change
more predictable.

\subsection*{Acknowledgments}

The author gratefully thanks Prof. Arthur Rörsch, Prof. Peter A. Ziegler
and anonymous reviewers for detailed suggestions.

\section*{Appendix: Willson and Hoyt's statements regarding the TSI satellite
records}

In 2008 the author inquired with Dr. Willson, who heads the ACRIM
satellite TSI measurements, and Dr. Hoyt (the inventor of GSN - Group
Sunspot Number indicator) who was in charge of the Nimbus7/ERB satellite
measurements, about their opinion regarding the theoretical modifications
applied to their published TSI records by Dr. Fröhlich of the PMOD/WRC
team. These modifications are crucial for obtaining a TSI satellite
composite record that does not show an increasing trend between 1980
and 2000 \citep{Frohlich1998,Frohlich}, as shown in Figure 14. For
detailed information visit ACRIM web-site.%
\footnote{\href{http://acrim.com/TSI\%5C\%20Monitoring.htm}{http://acrim.com/TSI\textbackslash{}\%{}20Monitoring.htm}%
}

In the statements reproduced in Figure 18, Willson and Hoyt agree
that Fröhlich's modifications are, in their opinion, not justified
because they are inconsistent with the physical properties of the
experimental instruments used for TSI satellite measurements. Of course,
these statements do not automatically imply that Fröhlich's modifications
are necessarily erroneous. However, it is clear that Willson and Hoyt,
who are the principal investigators of the experimental teams in charge
of the TSI satellite records modified by Fröhlich, are convinced that
the modification of their TSI records are not justified and that the
PMOD TSI satellite composite does not correspond to the actual TSI
satellite measurements and does not properly describe the actual dynamic
behavior of TSI from 1978 onward.

\end{document}